\documentclass[lettersize,journal]{IEEEtran}
\usepackage{amsmath,amsfonts}
\usepackage{algorithmic}
\usepackage{algorithm}
\usepackage{array}
\usepackage[caption=false,font=normalsize,labelfont=sf,textfont=sf]{subfig}
\usepackage{amsthm,cite,url,graphicx,booktabs,lipsum,color,bm,caption,soul}
\usepackage{textcomp}
\usepackage{stfloats}
\usepackage{url}
\usepackage{verbatim}
\usepackage{graphicx}
\usepackage{cite}
\usepackage{threeparttable}
\usepackage{booktabs}
\usepackage{multirow}
\usepackage{float}
\usepackage[utf8]{inputenc}
\usepackage[T1]{fontenc}

\usepackage{bm}
\usepackage{makecell} 
\begin{document}
	
	\title{QHARMA-GAN: Quasi-Harmonic Neural Vocoder based on Autoregressive Moving Average Model}
	
	\author{Shaowen Chen and 
		Tomoki Toda
		\IEEEcompsocitemizethanks{\IEEEcompsocthanksitem This work was partly supported by JST CREST Grant Number JPMJCR19A3.
			\IEEEcompsocthanksitem Shaowen Chen is with the Graduate School of Informatics, Nagoya University, Nagoya 464-8601, Japan (e-mail:shaowen.chen@g.sp.m.is.nagoya-u.ac.jp)
			\IEEEcompsocthanksitem Tomoki Toda is with the Information Technology Center, Nagoya University,
			Nagoya 464-8601, Japan (e-mail:tomoki@icts.nagoya-u.ac.jp)
		}
	}
	
	\markboth{Journal of \LaTeX\ Class Files,~Vol.~14, No.~8, August~2021}%
	{Shell \MakeLowercase{\textit{et al.}}: A Sample Article Using IEEEtran.cls for IEEE Journals}
	
	\IEEEaftertitletext{\begin{center}
			{\textit{This manuscript is currently under review for publication in the IEEE Transactions on Audio, Speech, and Language Processing.
					This work has been submitted to the IEEE for possible publication. Copyright may be transferred without notice, after which this version may no longer be accessible.}}
	\end{center}}

	\maketitle
	

	\begin{abstract}
		Vocoders, encoding speech signals into acoustic features and allowing for speech signal reconstruction from them, have been studied for decades. Recently, the rise of deep learning has particularly driven the development of neural vocoders to generate high-quality speech signals. On the other hand, the existing end-to-end neural vocoders suffer from a black-box nature that blinds the speech production mechanism and the intrinsic structure of speech, resulting in the ambiguity of separately modeling source excitation and resonance characteristics and the loss of flexibly synthesizing or modifying speech with high quality. Moreover, their sequence-wise waveform generation usually requires complicated networks, leading to substantial time consumption. In this work, inspired by the quasi-harmonic model (QHM) that represents speech as sparse components, we combine the neural network and QHM synthesis process to propose a novel framework for the neural vocoder. Accordingly, speech signals can be encoded into autoregressive moving average (ARMA) functions to model the resonance characteristics, yielding accurate estimates of the amplitudes and phases of quasi-harmonics at any frequency. Subsequently, the speech can be resynthesized and arbitrarily modified in terms of pitch shifting and time stretching with high quality, whereas the time consumption and network size decrease. The experiments indicate that the proposed method leverages the strengths of QHM, the ARMA model, and neural networks, leading to the outperformance of our methods over other methods in terms of generation speed, synthesis quality, and modification flexibility.
		
	\end{abstract}
	
	
	\section{Introduction}
	\IEEEPARstart{S}{peech} signal modeling, as a cornerstone of speech processing, has received significant attention and studies for its role in transforming speech signals into acoustic features and synthesizing them back into speech signals. 
	Vocoders, which enable these encoding and decoding processes, play a pivotal role in deepening our understanding of the essence and structure of speech during encoding and enabling speech synthesis and modification through the manipulation of extracted features during decoding. Historically, vocoders are categorized into a conventional vocoder based on conventional signal processing (CSP) and a neural vocoder based on deep learning.

	Conventional vocoders, well-established by CSP algorithms, initially dominate speech modeling. For instance, WORLD, as a source-filter-based method, decomposes speech signals into source excitation signals and resonance filter parameters by extracting $f_0$ \cite{morise2009fast}, spectral envelope \cite{morise2015cheaptrick}, \cite{morise2015error}, and aperiodic parameter \cite{morise2016d4c}. This decomposition allows for fast speech synthesis and modifications, such as time-scale or pitch-scale modifications, leading to good controllability; nevertheless, it suffers from fast yet inaccurate feature extraction, resulting in degraded synthesis quality. Moreover, WORLD's lack of phase modeling introduces discrepancies in generated waveforms compared with ground truth signals.

	As another approach to model speech with CSP, the spectral transform methods, including short-time Fourier transform (STFT) and its post-processing like synchrosqueezing transform (SST) \cite{2011Daubechies}, \cite{2013Thakur}, are extremely fast and convert the time-domain speech waveforms into spectrograms, allowing for the visualization of the speech signals' structure. Speech signals are typically nonstationary, are composed of deterministic harmonics and stochastic noise, and can be formulated as
	\begin{align}
	x(t)=\sum_{k=-K}^{K}x_k(t)+\varepsilon(t)=\sum_{k=-K}^{K}A_k(t)e^{i\varphi_k(t)} + \varepsilon(t),
	\label{eq1}
	\end{align}
	where $K$ is the number of harmonics, $A_k(t)$ and $\varphi_k(t)$ respectively denote the instantaneous amplitude and instantaneous phase of the $k$-th component at instantaneous time $t$, and $\varepsilon (t)$ is stochastic noise. These features reflect information such as the loudness, timbre, and pitch (also referred to as $f_0$) of the speech, implying that modifying these features enables the modification of the speech signal. However, the features are usually inaccurately extracted by STFT and SST, resulting in the degradation of speech resynthesis and modification. Hence, many time--frequency analysis methods, such as higher-order-SST \cite{oberlin2015second}, \cite{pham2017high}, MSST \cite{wang2013matching}, Stat-SST \cite{chen2023instantaneous}, and SET \cite{yu2017synchroextracting}, have been proposed to more accurately extract these acoustic features. However, they struggle to accurately extract unvoiced speech features, as these are not sparse in the frequency domain but instead spread across the entire frequency band.

	In contrast, the quasi-harmonic model (QHM) \cite{pantazis2008properties} and its derivatives, e.g., adaptive QHM (aQHM) \cite{pantazis2010adaptive} and the extension of aQHM (eaQHM) \cite{kafentzis2012extension}, have been proposed to model speech waveforms (including voiced and unvoiced parts) as the complex amplitude and frequency of sparse quasi-harmonic components, with which the speech can be almost perfectly reconstructed and modified. However, such high-quality speech synthesis achieved by QHM methods relies on iterative complex amplitude extractions and frequency corrections for each component, making the process computationally intensive and unsuitable for real-time applications. Moreover, QHM methods lack complex spectral envelope modeling, limiting their accuracy in parameter estimation during pitch-scale speech modification, consequently degrading speech quality. Given the above, conventional vocoders usually struggle to balance high speed and high quality while being insufficiently robust, leading to their vulnerability to noise disturbances.

	On the other hand, emerging deep neural networks (DNNs) have hastened the development of neural vocoders by leveraging their powerful fitting capabilities through deeper architectures, establishing strong and complicated mappings between input speech and output speech. In early advancements, such as WaveNet, an autoregressive framework was utilized for pointwise waveform generation, pioneering neural vocoder development. However, its computation cost is heavy. Thus, Parallel WaveNet \cite{oord2018parallel} with knowledge distillation was proposed to accelerate the generation while maintaining speech quality. WaveGlow \cite{prenger2019waveglow} combined Glow and WaveNet into a flow-based model, achieving a balance between speed and quality, whereas WaveRNN \cite{kalchbrenner2018efficient} adopted the recurrent neural network (RNN) to significantly reduce the computation cost and enhance speech quality. 
	Afterward, the emergence of generative adversarial networks (GANs) catalyzed the development of GAN-based methods. GAN-based models, such as Parallel WaveGAN \cite{yamamoto2020parallel} and MelGan \cite{kumar2019melgan}, \cite{mustafa2021stylemelgan}, introduced a competitive framework between a generator and a discriminator, driving the generator to produce outputs indistinguishable from ground truth. As a widely used GAN-based vocoder, HiFi-GAN \cite{kong2020hifi} achieved high-speed speech synthesis with impressive quality by converting the mel-spectrogram to speech waveforms through multiple upsamplings in the time domain and compressions in the frequency domain. Because of the sophisticated design, these models demonstrate robustness and high-quality generation, provided they are trained on sufficiently large datasets. Namely, they are data-hungry and require extensive amounts of training data to avoid overfitting.

	Despite the existing advancements in speech modeling, vocoders are increasingly challenged by the need for real-time speech generation \cite{quatieri1986speech} and speech modification \cite{ninness2008time}, \cite{quatieri1992shape} with high quality. Although neural vocoders can generate higher-quality speech than conventional vocoders, their complex structures often result in a slow inference. Additionally, most neural vocoders integrate both encoding and decoding processes, directly converting the input into speech signals without extracting interpretable, physically meaningful parameters from the speech signal, such as the resonance characteristics and fundamental frequency ($f_0$). Such processing makes it impossible to directly and reliably modify speech. LPCNet \cite{valin2019lpcnet} was proposed to provide excitation signals via DNN and model the vocal tract by linear prediction (LP), on the basis of which the speech from specific speakers can be synthesized. However, as observed in QHM methods using the discrete all-pole (DAP) \cite{el1991discrete}, the LP coefficients estimated by the Levinson--Durbin algorithm for LPCNet are prone to inaccuracies, often degrading vocal tract modeling. Consequently, vocoders, such as SiFi-GAN \cite{yoneyama2023source} and HN-uSFGAN \cite{yoneyama2022hu}, adopt source-filter architectures to model excitation signals and vocal tracts by QP-ResBlocks \cite{wu2021quasiwavenet}, \cite{wu2021quasiwavgan}, enabling $f_0$ modification. 
	However, their performance is limited and the structures are usually complicated, increasing the burden of computation.

	To tackle the issues of existing neural vocoders, including slow generation and the inability to reveal signal structures, which hinders high-quality speech modification, we propose a novel framework for vocoder by combining CSP and DNN, thereby visualizing the speech structure and enabling the flexible speech modification.
    In our previous work, 
	QHM and DNN were successfully integrated to develop a novel vocoder, i.e., QHM-GAN \cite{chen20qhmgan}. This allows redundant speech signals to be represented as interpretable parameters of sparse frequency components, implying that the speech can be modified with high quality by manipulating the sparse parameters. 
	To achieve speech modeling and speech modification via $f_0$ control,
	in this paper, we extend our previous work by embedding the autoregressive moving average (ARMA) model into QHM-GAN, enhancing the resonance characteristic modeling using DNN. As a result, the proposed vocoder demonstrates improvements in synthesis quality, inference speed, and pitch controllability. The main contributions of this paper are summarized as follows:
	\begin{itemize} 
		\item [ 1)] To simultaneously achieve the controllability and fast generation of CSP, as well as the robustness and high-quality generation of DNN, we combine QHM and DNN into a novel framework for vocoders. Compared with QHM, we simplify the instantaneous phase generation by defining phase compensation to accelerate the synthesis process while maintaining the frequency correction mechanism. By using the parameters estimated by DNN, the speech can be rapidly resynthesized with high quality. This contribution is identical to that in our previous work \cite{chen20qhmgan}, and is included here for completeness.
		\item [ 2)] To achieve the ability to explicitly model the resonance characteristics, we incorporate the ARMA model for QHM-GAN, where ARMA parameters are estimated by DNN, making it possible to obtain the amplitude and phase delay of an arbitrary quasi-harmonic component from the frequency response of the ARMA model, leading to more efficient speech modeling and flexible speech modification.
		\item [ 3)] On the basis of the proposed vocoder, a speech modification algorithm for the time stretching and pitch scaling of speech is introduced, ensuring that the shape of the spectral envelope (vocal tract response) remains unchanged before and after the modification.
		
	\end{itemize} 
	
	The structure of this paper is outlined as follows. In Section II, we review QHM methods and HiFi-GAN, discussing their limitations. In Section III, the ideas and structure of the proposed vocoder are introduced. In Section IV, the inspiration, derivation, and details of resonance modeling based on the proposed vocoder for speech modification are elaborated. In Section V, the quantitative performance is compared between our proposed method and other methods by analyzing speech segments. The paper concludes with a summary in Section VI.
	
	\section{Related Works}
	\subsection{QHM Methods}
	QHM methods work on the basis of the assumption that each frame of signals can be decomposed into several sinewaves with different frequencies:
	\begin{align}
	x(t)=\left( \sum\limits_{k=-K}^{K}{{{a}_{k}}{{e}^{j2\pi {{f}_{k}}t}}} \right)w(t),\text{   }t\in \left[ -{{T}_{l}},{{T}_{l}} \right],
	\label{eq2}
	\end{align}
	
	\noindent where $a_k$ and $f_k$ are the complex amplitude and frequency of the $k$-th component, respectively. $w(\cdot)$ means the window with the length $2T_l$. The frequency $f_k$ is not strictly $k$ times the pitch $f_0$, i.e., $f_k \neq kf_0 $, hence, these methods are referred to as quasi-harmonic model methods. Because of the ubiquitous deviation of $f_0$ extracted using a pitch detector from the true value, QHM methods use complex linear functions to fit the frequency mismatch and simultaneously obtain the complex amplitude using the following framewise models:
	\begin{subequations}
		\begin{align}
		\text{QHM: }x(t)&=\left[ \sum\limits_{k=-K}^{K}{\left( {{a}_{k}}+t{{b}_{k}} \right){{e}^{j2\pi {{{\hat{f}}}_{k}}t}}} \right]w(t), \hfill \label{eq2_1} \\ 
		\text{aQHM: }x(t)&=\left[ \sum\limits_{k=-K}^{K}{\left( {{a}_{k}}+t{{b}_{k}} \right){{e}^{j\varPhi_k(t)}}} \right]w(t),\hfill \label{eq2_2}\\
		\text{eaQHM: }x(t)&=\left[ \sum\limits_{k=-K}^{K}{\left( {{a}_{k}}+t{{b}_{k}} \right)\frac{{{A}_{k}}(t+{{t}_{l}})}{{{A}_{k}}({{t}_{l}})}{{e}^{j\varPhi_k(t)}}} \right]w(t),\hfill \label{eq2_3}\\
		\text{and } \varPhi_k(t) &=  {{{\hat{\varphi }}}_{k}}(t+{{t}_{l}})-{{{\hat{\varphi }}}_{k}}({{t}_{l}})  \text{ and }t\in \left[ -{{T}_{l}},{{T}_{l}} \right], \hfill \notag
		\end{align}
	\end{subequations}
	\noindent where ${{\hat{f}}_{k}}$ and ${{b}_{k}}$ denote the initially estimated frequency and complex slope of the $k$-th component, respectively. $A_k(t)$ and ${\hat{\varphi}_k}(t)$ denote the amplitude and phase function of the $k$-th harmonic component, respectively, and $t_l$ is the center of the $l$-th frame ($l=1,\cdots,L$). 
	${{b}_{k}}$ enables the model to adaptively match the speech waveforms, compensating for mismatch between original speech and synthetic speech when the initial frequency is inaccurate, so that the model has a frequency correction mechanism. First, $a_k$ and $b_k$ are treated as variables and obtained by solving Eqs. (\ref{eq2_1})--(\ref{eq2_3}) via least squares (LS) optimization between models and target speech. Then, the frequency can be corrected as
	\begin{align}
	{{f}_{k}}={{\hat{f}}_{k}}+{{\eta }_{k}}={{\hat{f}}_{k}}+\frac{1}{2\pi }\frac{a_{k}^{R}b_{k}^{I}-a_{k}^{I}b_{k}^{R}}{{{\left| {{a}_{k}} \right|}^{2}}},
	\label{eq4}
	\end{align}
	\noindent where $a_{k}^{R},a_{k}^{I}$ and $b_{k}^{R},b_{k}^{I}$ denote the real and imaginary parts of ${{a}_{k}},{{b}_{k}}$, respectively. QHM assumes that the signal is stationary within each frame, which does not follow the facts, leading to inaccurate solutions for $a_k$ and $b_k$. To address this issue, aQHM imports a nonstationary phase function $\varPhi_k(t)$ for modeling, which better aligns with the time-varying frequency of speech signals. On the basis of aQHM, eaQHM adds an amplitude amplifier at the amplitude part of the model to approximate the real waveform whose amplitude varies rapidly over time, as shown in Eq. (\ref{eq2_3}). As models increasingly gain the ability to capture the nonstationarity of both amplitude and phase, the amplitude and phase can be extracted with greater accuracy, ultimately resulting in higher-quality speech reconstruction.
	
	Unfortunately, the increasing complexity of the models also makes the analysis processes increasingly time-consuming. 
	Since aQHM and eaQHM require iterations to gradually approach the true values. They initially use QHM to update the frequency and interpolate it to an instantaneous version $f_k(t)$ for calculating the instantaneous phase at each frame by
	\begin{align}
	{\varphi }_k(t)={\varphi }_k(t_l) + \int_{{t_l}}^{t_l+t}{2\pi {{{{f}}}_{k}}(u)}\text{d}u, \text{ }t\in [ -{{T}_{l}},{{T}_{l}}].\label{eq5}
	\end{align}
	Secondly, the instantaneous amplitude is calculated by linear interpolation to obtain the amplitude amplifier, whereas the nonstationary phase function subtract the old one for the new extraction of $a_k$ and $b_k$ in the next iteration. The frequency will gradually approach its true value until the synthetic speech stops approaching the original speech. Eventually, the entire speech, including unvoiced segments, can be effectively modeled and accurately resynthesized using the complex amplitude and frequency of quasi-harmonic components alone. 
	
	In synthesis, the framewise amplitude and phase, i.e., 
	\begin{align}
	{{{{A}}}_{k}}({{t}_{l}})=\left| {{{{a}}}_{k}}({{t}_{l}}) \right|, 
	{{{{\varphi }}}_{k}}({{t}_{l}})=\angle {{{{a}}}_{k}}({{t}_{l}}),
	\label{eq6}
	\end{align}
	will be interpolated to instantaneous versions. In particular, the instantaneous phase is computed in a special way: 
	\begin{align}
	{{\hat{\varphi }}_{k}}(t)={{\hat{\varphi }}_{k}}({{t}_{l}})+\int_{{{t}_{l}}}^{t_l+t}{2\pi {{{\hat{f}}}_{k}}(u)}+z\sin \left[ \frac{\pi (u-{{t}_{l-1}})}{{{t}_{l}}-{{t}_{l-1}}} \right]\text{d}u,
	\label{eq7}
	\end{align}
	where ${{\hat{\varphi }}_{k}}({{t}_{l}})$ is the current frame's phase obtained by Eq. (\ref{eq6}) and $z$ is computed by
	\begin{align*}
	z=\frac{\pi [{{{\hat{\varphi }}}_{k}}({{t}_{l+1}})+2\pi M-{{{\tilde{\varphi }}}_{k}}({{t}_{l+1}})]}{2({{t}_{l+1}}-{{t}_{l}})}.
	\end{align*}
	$\tilde{\varphi }_{k}({{t}_{l+1}})$ is obtained by Eq. (\ref{eq5}) and $M$ is the closest integer to ${\left| {{{\hat{\varphi }}}_{k}}(t_{l+1})-{{{\tilde{\varphi }}}_{k}}(t_{l+1}) \right|}/{2\pi }]$. Since it is difficult to guarantee that the phase obtained by LS optimization is equal to that obtained by integration, i.e., ${{{\tilde{\varphi }}}_{k}}(t_{l+1})= {{{\hat{\varphi }}}_{k}}(t_{l+1})+{2\pi}M$, accordingly, Eq. (\ref{eq7}) plays a vital role in smoothing the phase, which is essential for achieving high-quality speech synthesis.
	
	QHM methods often need a sufficiently wide window to ensure the stability of LS. For instance, from Eqs. (\ref{eq2_1})--(\ref{eq2_3}), at least $4K$ samples are required to ensure the stable LS solutions for $a_k$ and $b_k$. When $K$ is large, the window becomes wider, resulting in an increase in computational complexity.
	Moreover, the framewise extraction of QHM methods ignores information from adjacent frames, resulting in discontinuities in the extracted parameters and consequently degrading the synthesis quality; such degradation becomes particularly pronounced when the errors of the initial $f_0$ are substantial.
	
	To address the aforementioned issues, a sequence-wise method, named BP-QHM, is proposed, which backpropagates the speech waveform loss to the QHM parameters through the synthesis process, subsequently optimizing the parameters through gradient descent \cite{bpqhm}. By considering adjacent frames, the extracted parameters are rendered more accurate and continuous. However, since backpropagation requires multiple iterations, the time consumption is still substantial. The positive aspect is that the success of BP-QHM also highlights the potential of QHM application in deep learning.
	
	\subsection{HiFi-GAN}
	HiFi-GAN employs a sophisticatedly designed multi-scale and multi-period discriminator. The generator gradually upsamples features using several transposed convolution layers, aligning the resolution of output with that of the target speech. After each transposed convolution, the intermediate features are passed through a multi-receptive field fusion (MRF) module, which comprises several residual blocks. Each block applies multiple dilation sizes to capture the input at different scales. As the generator endeavors to deceive the discriminator, the discriminator is trained to distinguish between generated speech and real speech. The discriminator consists of a multi-period discriminator and a multi-scale discriminator, which distinguishes the authenticity of speech from the multi-period and multi-scale perspectives. During the training, the generator is optimized using the adversarial loss $L_\text{g,adv}$, feature matching loss $L_\text{fm}$, and mel-spectrogram loss $L_\text{mel}$:
	\begin{align}
	L_\text{G} = L_\text{g,adv} + \lambda_\text{fm}L_\text{fm} + \lambda_\text{mel}L_\text{mel},
	\label{eq8}
	\end{align}
	where $\lambda_\text{fm}$ and $\lambda_\text{mel}$ are weight coefficients for feature matching loss $L_\text{fm}$ and mel-spectrogram loss $L_\text{mel}$, respectively. $L_\text{fm}$ and $L_\text{mel}$ adopt the $l^1$ norm to measure the distance between the hidden feature and mel-spectrograms of the generated speech and ground truth, respectively. On the contrary, the discriminator is updated by only the adversarial loss $L_\text{d,adv}$, which is composed of the losses from different period and scale discriminators.

	The sophisticated structure of the generator makes the generation process time-consuming. While HiFi-GAN is among the faster neural vocoders, the use of several transposed convolutions hampers its applicability to real-time processing. Additionally, many studies, such as \cite{yoneyama2023source},\cite{yoneyama2024wavehax},\cite{matsubara2023harmonic}, proves that HiFi-GAN struggles to interpret the intrinsic structure of speech, thus limiting its ability to modify features and perform speech scaling. Without the absorption of $f_0$ priors, HiFi-GAN cannot perform $f_0$ extrapolation. Although a modified HiFi-GAN includes $f_0$ input to control $f_0$ \cite{matsubara2023harmonic}, \cite{HiFiGANf0}, \cite{HiFiGANf02}, this control is restricted to the range observed in training data, making it challenging to modify $f_0$ beyond this range. Moreover, HiFi-GAN's inability to model the speech signal structure leads to high data dependence, requiring a large dataset to avoid overfitting. As a result, its generalization capability is limited, leading to significant data cost for effective training.

	\section{QHM-GAN: Neural Vocoder based on Quasi-Harmonic Modeling}
\begin{figure*}[!t]\centering
		\includegraphics[width=17cm]{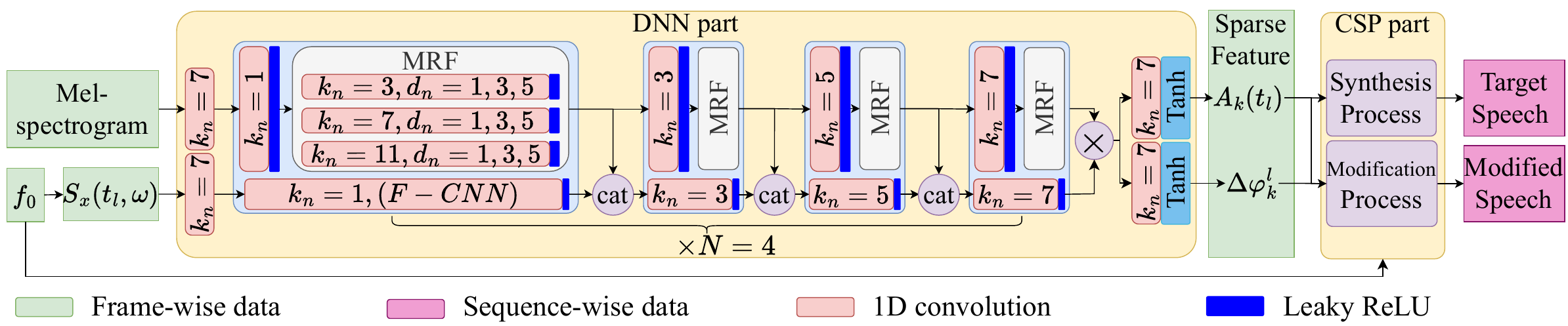}
		\vspace{-0.2cm}
		\caption{Structure of QHM-GAN generator, which takes mel-spectrogram as inputs and outputs sparse framewise amplitude and phase compensation. $k_n$ and $d_n$ are the kernel size and dilation of the corresponding convolution layer, respectively, while Tanh denotes the hyperbolic tangent function. If unlabeled, $d_n=1$ by default. Note that the configurations of all MRFs in the generator are the same.}
		\label{QHM_GAN_structure}
		\vspace{-0.3cm}
	\end{figure*}

	To simultaneously address the limitations of conventional and neural vocoders, we aim for a vocoder that efficiently and accurately compresses speech signals into sparse features during encoding, facilitating transmission and subsequent processing, as the speech signals have large data volumes owing to the high sampling rate. During decoding, the vocoder must rapidly and reliably reconstruct high-quality speech signals.
	Inspired by the ability of QHM methods to sparsely model speech signals (both voiced speech and unvoiced speech), generate speech efficiently, and provide precise $f_0$ control, combined with the robust inference capabilities of DNNs, we propose a novel vocoder framework that integrates CSP and DNN, named QHM-GAN, to leverage their advantages. Compared with the mel-spectrogram-to-speech structure, QHM-GAN adopts a mel-spectrogram-to-parameter structure to accurately generate the quasi-harmonic parameters, with which the speech signals can be efficiently resynthesized and modified, allowing for $f_0$ extrapolation. The generator's structure is shown in Fig. \ref{QHM_GAN_structure}.
	\subsection{QHM-GAN}
	QHM-GAN consists of two parts: the DNN part and CSP part. The DNN part takes the mel-spectrogram as inputs and outputs sparse quasi-harmonic parameters, i.e., the framewise amplitude $A_k(t_l)$ and phase compensation $\Delta \varphi_k^l$. On the other hand, the CSP part handles decoding, using the sparse parameters encoded by the DNN part to generate speech. Below, these two parts will be introduced separately.
	\subsubsection{Synthesis process with phase compensation}
	In the synthesis process, the speech is generated from the framewise parameters, i.e., the pitch $f_0$, amplitude, and phase, which analytically represent the structure of speech waveforms. Once the instantaneous versions of amplitude and phase are obtained, the speech can be resynthesized by 
	\begin{align}
	\hat{x}(t) = \sum_{k=-K}^{K}\hat{A}_k(t)e^{i\hat{\varphi}_k(t)},
	\label{eq9}
	\end{align}
	where $\hat{(\cdot)}$ means the estimate of the corresponding variable. The instantaneous phase is usually calculated by Eq. (\ref{eq7}); nevertheless, the prerequisite is that the error between $\tilde{\varphi }_{k}({{t}_{l+1}})$ and unwrapped $\hat{\varphi }_{k}({{t}_{l+1}})$ is small. The frequencies estimated by pitch detectors usually deviate from their true value, and a large deviation brings about a great error between $\tilde{\varphi }_{k}({{t}_{l+1}})$ and unwrapped $\hat{\varphi }_{k}({{t}_{l+1}})$. Additionally, it is well known that unvoiced speech signals are not harmonical, thus, using harmonics to match unvoiced speech signals inevitably results in larger errors. Given the above, a method to compensate for the phase error, similar to QHM methods, is essential.
	
	To replicate the ability of QHM methods to adaptively adjust the frequency, we enable DNNs to estimate the phase compensation for each component at the center of each frame and denote it as $\Delta \varphi_k^l$. Subsequently, after obtaining the individual instantaneous frequency by $\hat{f}_k(t) = k\hat{f}_0(t)$, the phase at the center of each frame can be corrected as   
	\begin{align}
	\hat{\varphi }_k(t_l)= \int_{0}^{t_l}{2\pi {\hat{f}_{k}}(u)} \text{d}u + \sum_{i=0}^{l}\Delta\varphi_k^i.
	\label{eq10}
	\end{align}
	Subsequently, we employ DNNs to estimate this discrepancy in order to compensate for the phase at each frame center. For ease of calculation, we use the average frequency between frame centers to linearly compute the phase difference as
	\begin{align}
	\int_{0}^{t_{l}}{2\pi {\hat{f}_{k}}(u)} \text{d}u \approx \pi\sum_{i=1}^{l}{[\hat{f}_k^{i-1} + \hat{f}_k^{i}] (t_{i} - t_{i-1})},
	\label{eq11}
	\end{align}
	where $\hat{f}_k^l$ means the estimated frequency of the $k$-th component at the $l$-th frame.
	Then, the instantaneous phase can be depicted by cubic interpolation, which is used for speech synthesis along with the instantaneous amplitude linearly interpolated by the framewise version. By introducing phase compensation, the generator gains the ability to autonomously adjust frequencies, similar to QHM methods. This enhancement enables the generated speech to better match the target speech waveform in terms of both voiced and unvoiced speech signals.
	
	\subsubsection{Framewise Parameters Estimated by DNN}
	The goal of the DNN part is to estimate the framewise amplitude and phase compensation from the mel-spectrogram for further resynthesis. Given that speech is a temporal sequence, temporal DNNs, such as RNNs, LSTMs, or CNNs, are preferred for their ability to consider both past and future contexts in the input. Since QHM-GAN needs to determine the parameters of the corresponding harmonics on the basis of input $f_0$, the generator must not only absorb the mel-spectrogram but also $f_0$. Therefore, we directly employ MRF modules to construct the main part of the network to capture the linguistic and speaker information. Additionally, we utilize CNNs to process $f_0$ to screen out the desirable hidden features from the result of MRF modules, which helps control the pitch. Because the input and output of the generator are both framewise, we use normal CNN layers to connect each MRF. All activation functions between blocks are leaky ReLUs.

	Here, we detail how the mel-spectrogram and $f_0$ are fed into and processed by the network. First, the mel-spectrogram is fed through a normal CNN layer. The result is then sequentially passed through $N$ MRF modules (typically $N=4$) to extract hidden features. On the other hand, to input $f_0$ into the network in a spectrogram-like form, we generate a pseudo-STFT where the amplitudes of all components are 1, as
	\begin{align}
	{S}_x(t_l,\omega)= \sum^{K}_{k=-K} e^{-\frac{\sigma^2 [\omega - 2\pi \hat{f}_k(t_l)]^2}{2}},
	\label{eq12}
	\end{align}
	where $\sigma$ is the standard deviation of the Gaussian window. This is because when the frequency and amplitude of a signal are weakly modulated within a frame, its STFT can be approximated as a superposition of window functions located at harmonic frequencies. Then, the pseudo-STFT will pass through several normal CNNs (called F-CNNs), and finally, a hidden feature containing the harmonical frequency information is obtained. Note that after each F-CNN, the result is concatenated to the corresponding MRF result and used as input for the next F-CNN. After passing through several F-CNNs, the final result will multiply the hidden feature from the final MRF module to highlight the desirable information, which will be further separately fed into two normal CNN blocks to generate the phase compensation and amplitude. Since the phase compensation is naturally constrained within $[-\pi, \pi]$, we employ the hyperbolic tangent function to constrain the output result within the domain. The generated amplitude will be interpolated to the instantaneous amplitude, whereas the generated phase compensation will be used to obtain the framewise phase by Eq. (\ref{eq10}), which will be cubically interpolated to the instantaneous phase. Eventually, the speech will be synthesized using these instantaneous products by Eq. (\ref{eq9}). The pseudocode of speech synthesis by QHM-GAN is elaborated in Algorithm \ref{alg1}.
	The differentiability of the entire framework facilitates the backpropagation of gradients from the loss function. Moreover, it is the phase compensation that enables the model to correct the phase at each frame and thus adjust the frequency of the sparse components.
	
	\begin{algorithm}
		\caption{Speech synthesis based on QHM-GAN.}
		\begin{algorithmic}
			\STATE {\textbf{Step 1: Preprocessing and Setting}\\
				Extract the $\hat{f}_k$ and mel-spectrogram from speech $x(t)$ and compute $S_x(t,\omega)$;
				set sampling rate $f_s$, harmonic number $K$, and frame-shift; 
				obtain $\hat{A}_k(t_l)$ and $\Delta \varphi_k^l$ from DNN;}
			
			\STATE {\textbf{Step 2: Instantaneous amplitude and phase}\\
				Compute raw rotation angle by Eq. (\ref{eq11});\\
				Compute unwrapped framewise phase $\hat{\varphi}_k(t_l)$ by Eq. (\ref{eq10});\\
				Cubically interpolate $\hat{\varphi}_k(t_l)$ into $\hat{\varphi}_k(t)$;\\
				Linearly interpolate $\hat{A}_k(t_l)$ into $\hat{A}_k(t)$;\\
			}
			\STATE {\textbf{Step 3: Generation}\\
				\STATE $\hat{x}(t) \leftarrow \sum_{k=-K}^{K}\hat{A}_k(t)e^{i\hat{\varphi}_k(t)}$;
			}
			\STATE {\textbf{Output:} $\hat{x}(t)$}
		\end{algorithmic}
		\label{alg1}
		\vspace{-0.1cm}
	\end{algorithm}
	
	\subsubsection{Training}
	The training process of QHM-GAN is based on that of HiFi-GAN. However, the human ear is sensitive to frequencies instead of time-domain waveforms. Therefore, we adopt the spectrogram-based discriminator described in \cite{UnivNet} to distinguish the result from the target speech in terms of the multi-resolution of the spectrogram. Such a spectrogram-based discriminator will act as a substitute for the human ear in discerning frequency variations within speech, thereby encouraging the generator to more effectively learn to produce speech with smooth frequency variations akin to the target speech.
	
	\subsection{High-speed Generation}
	Since QHM-GAN does not necessitate upsampling, it avoids the time-consuming transposed convolution. Furthermore, the generator yields outputs with the same resolution as the mel-spectrograms, leading to the data convolved by the kernels not increasing in size owing to upsampling, undoubtedly preventing an increase in computational time. Therefore, QHM-GAN achieves rapid inference.
	Moreover, compared with methods like HiFi-GAN that convert the mel-spectrogram to a speech waveform, the DNN part of QHM-GAN generates the sparse spectral parameters (amplitude and phase), significantly reducing the learning burden on the network. Hence, we can simplify the generator by reducing the number of MRF modules or the number of dilations in each MRF's kernel while maintaining accurate inference capabilities, further increasing the generation speed. The result of our preliminary experiments demonstrate that reducing the number of MRF modules and decreasing the number of dilations do not significantly compromise the quality of generated speech, implying the viability of applying QHM-GAN in real-time processing.
	
	\subsection{Inaccurate Speech Modification}
	Speech modification is defined as the change in the $f_0$ or duration of speech while preserving the short-time spectral envelope characteristics of speech (also regarded as the resonance characteristics), i.e., time- and pitch-scale modifications, which are closely related to the amplitude, frequency, and phase of speech. QHM-GAN is well suited for this task. For time-scale modification, we only need to stretch or shrink the framewise parameters obtained from QHM-GAN over time, i.e., interpolate framewise parameters with modified frame-shift. This process preserves the spectral envelope, thereby achieving the effective stretching of the speech waveform without altering its spectral characteristics. On the other hand, regarding pitch-scale modification, QHM-GAN lacks the capability to model the spectral envelope, which limits its performance in the accurate estimation of the parameters for modified frequencies. Our informal experiments unfortunately show that the amplitude estimated by directly feeding the modified $f_0$ into the DNN part of QHM-GAN is inaccurate, since there is no reference for a modified speech during training, making it challenging for DNN to learn the mapping accurately. Thus, we need an additional estimation of the spectral envelope based on the amplitude obtained by QHM-GAN to keep it fixed during modification. Conventional methods, such as LPC and DAP, are usually chosen to estimate the spectral envelope. However, their unrobust and unsatisfactory results degrade the synthesis quality, ultimately diminishing the effectiveness of speech modification.

	\section{QHARMA-GAN: Neural Vocoder based on QHM and ARMA Modeling}
	
	To overcome the limitation of QHM-GAN that fails to model the spectral envelope, we enhance the generator of QHM-GAN by embedding an ARMA model, enabling it to learn the spectral envelope effectively. In this section, the enhanced QHM-GAN and its application in speech modification are comprehensively discussed.
	
	\subsection{ARMA Modeling}
	LPC is renowned for its proficiency in analyzing discrete time-series signals, such as speech signals, by inferring the present value of a signal by considering the previous values. As a powerful model in excavating the autocorrelation structure of signals, ARMA model combines the previous input and outputs signals to predict the present value of the output signal as
	\begin{align}
	x(t)=-\sum_{p=1}^{P}a_px(t-p) + G\sum_{q=0}^Qb_qu(t-q),
	\label{eq13}
	\end{align} 
	where $b_0=1$ and $G$ is the gain. $x(t)$ and $u(t)$ are the output and input signals, whereas $P$ and $Q$ are the orders of autoregressive (AR) and moving average (MA) models, respectively. $a_p$ and $b_q$ are the AR and MA coefficients, respectively. Since speech is a typical time-series signal, ARMA can be employed to excavate the structure of speech. By considering the output signal as the desirable speech and the input signal as the excitation signal from the glottis, the ARMA model can be viewed as the resonance filter that process the excitation signal. The inherent frequency response of the ARMA  model will attenuate the energy and delay the phase of each frequency differently, producing a specific speech waveform. 
	Denoting the ARMA coefficients for the $l$-th frame as $G_l$, $a^l_p$, and $b^l_q$, we derive the process of ARMA modeling at the $l$-th frame.

	First, we assume that the excitation signal is composed of harmonics, each with an amplitude of 1, as 
	\begin{align}
	u^l(t)=\sum_{k=-K}^{K}u^l_k(t)=\sum_{k=-K}^{K}e^{i2\pi \hat{f}^l_kt}, t\in \left[ -{{T}_{l}},{{T}_{l}} \right],
	\label{eq14}
	\end{align}
	where $\hat{f}^l_k$ is given by the pitch detector.
	Considering the ARMA model to be time-varying for the entire speech signal, we denote it as $h(t)$ in the time domain and $H(t,\omega)$ in the time--frequency domain. Thus, $H(t,\omega)$ at the $l$-th frame reads 
	\begin{align}
	H(t_l,\omega) = \frac{X(t_l,\omega)}{U(t_l,\omega)}=G_l\frac{1+\sum_{q=1}^Qb^l_qe^{-i\omega q}}{1+\sum_{p=1}^Pa^l_pe^{-i\omega p}},
	\label{eq15}
	\end{align}
	where $X(t_l,\omega)$ and $U(t_l,\omega)$ are the Fourier transforms of $x^l(t)$ and $u^l(t)$. Focusing on the $k$-th component, the amplitude and phase of $u^l_k(t)$ will be attenuated and delayed by the ARMA model, respectively. Thus, for the $k$-th component at the $l$-th frame, the amplitude of $x^l(t)$ can be calculated by
	\begin{align}
	\hat{A}_k(t_l) =|H(t_l,\omega)|= |G_l|\frac{|1+\sum_{q=1}^Qb^l_qe^{-i\omega_k q}|}{|1+\sum_{p=1}^Pa^l_pe^{-i\omega_k p}|},
	\label{eq16}
	\end{align}
	while the phase will be determined by
	\begin{align}
	\hat{\varphi }_k(t_l) =  \varphi^u_{k}(t_l) + \angle H(t_l,\omega_k).
	\label{eq17}
	\end{align}
	$\varphi^u_{k}(t)$ is the phase of the excitation signal and is modeled by
	\begin{align}
	\varphi^u_{k}(t_l) = 	\int_{0}^{t_{l}}{2\pi {\hat{f}_{k}}(u)} \text{d}u,
	\label{eq18}
	\end{align}
	and $\hat{f}_k(u)$ is the instantaneous frequency interpolated from the framewise frequency $\hat{f}^l_k$. Evidently, once the ARMA coefficients are obtained, the amplitude and phase of each frequency component can be determined using the aforementioned equations. Consequently, when modifying the $f_0$ of speech, we can flexibly compute the corresponding amplitude and phase of the modified frequencies, enabling the rapid and hiqh-quality generation of the modified speech.
	To achieve this goal, we are inspired by QHM-GAN to employ DNNs, which contain a 1-channel output layer, a $P$-channel output layer, and a $Q$-channel output layer, to estimate the parameters $G_l$, $a^l_p$, and $b^l_q$ for each frame, respectively. The determined ARMA models are used to compute $\hat{A}_k(t_l)$ and $\hat{\varphi}_{k}(t_l)$ for generating speech signals.
	
	\subsection{QHARMA-GAN}
	Looking back to QHM-GAN, the phase compensation is generated to compensate for the error between the true and estimated rotation angles of each frame. Phase compensation is independent across frames and can be accumulated frame by frame during the frequency correction. Therefore, it is recommended to constrain the phase compensation generated by DNN within the range of $[-\pi,\pi]$ for a wide-band correction. In contrast, phase delay acts on the unwrapped phase in each frame, indicating that the phase delay in each frame affects the frequency of both the preceding and succeeding frames when correcting the frequency. Assuming a constant frequency within each frame, the frequency correction of the $k$-th component at the $l$-th frame ($\Delta f_k^l$) can be computed by
	\begin{align}
	\Delta f_k^l = \frac{ \hat{\varphi }_k(t_l) -  \hat{\varphi }_k(t_{l-1})  }{2\pi (t_{l}-t_{l-1})}  - \hat{f}_k^l .
	\label{eq19}
	\end{align}
	Denoting $t_{l}-t_{l-1} = \Delta t$ and $\Delta f_k^l$ based on phase delay as $\Delta \check{f}_{k}^l$, the cumulative $\Delta \check{f}_{k}^l$ for all frames can be computed by
	\begin{align}
	\sum_{l=1}^{L} \Delta \check{f}_{k}^l &= \sum_{l=1}^{L} \frac{\angle H(t_l,\omega_k) - \angle H(t_{l-1},\omega_k)}{2\pi \Delta t} \notag \\
	&= \frac{\angle H(t_L,\omega_k) - \angle H(t_{0},\omega_k)}{2\pi\Delta t} \in \left[\frac{-1}{\Delta t},\frac{1}{\Delta t} \right],
	\label{eq20}
	\end{align}
	where $\angle H(t_l,\omega_k) \in \left[-\pi,\pi\right]$ is considered. 
	Likewise, denoting $\Delta {f}_{k}^l$ based on phase compensation as $\Delta \acute{f}_{k}^l$, the cumulative  $\Delta \acute{f}_{k}^l$ for all frames can be computed by
	\begin{align}
	\sum_{l=1}^{L} \Delta \acute{f}_{k}^l = \frac{\sum_{l=1}^{L}\Delta \varphi^l_{k}}{2\pi\Delta t} \in \left[\frac{-L}{2\Delta t},\frac{L}{2\Delta t} \right].
	\label{eq21}
	\end{align}
	Comparing Eqs. (\ref{eq20}) and (\ref{eq21}), it is apparent that $L>2$ and the phase delay will lead to a limited frequency correction. Consequently, to preserve the QHM-GAN's capability of wide-band frequency correction, a larger phase delay is necessary, for example $[-r\pi,r\pi]$. Next, a trick will be introduced to obtain the large phase delays. 
	
\begin{figure*}[!t]\centering
		\includegraphics[width=17cm]{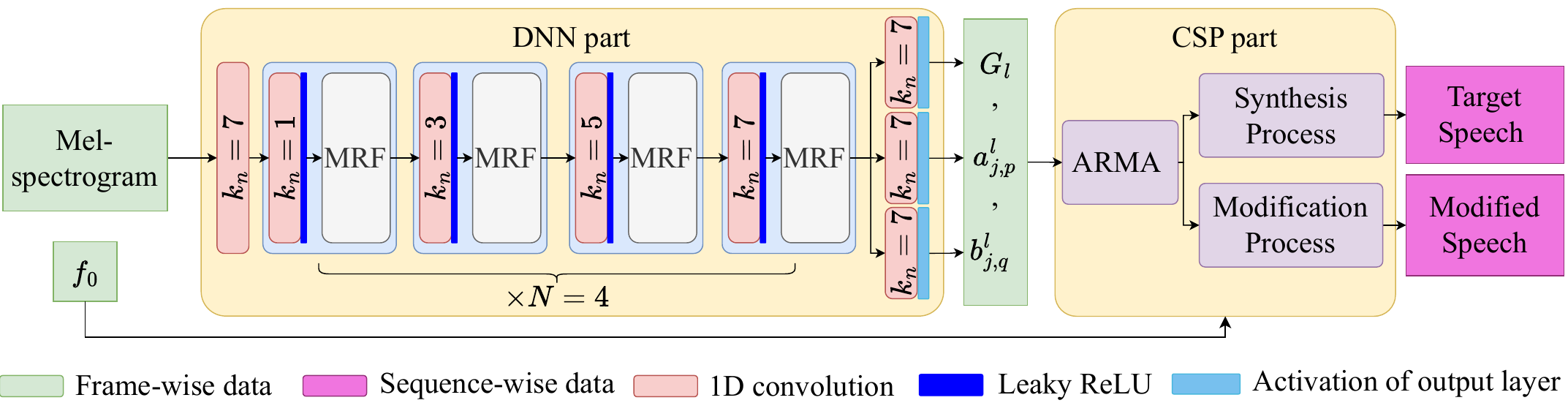}
		\vspace{-0.2cm}
		\caption{Structure of QHARMA-GAN generator, which takes mel-spectrogram as input and outputs ARMA coefficients. $k_n$ is the kernel size of the corresponding convolution layer. If unlabeled, a default dilation of 1 is used. The configurations of MRFs are the same as those shown in Fig. \ref{QHM_GAN_structure}. Note that the activation functions in the output layer can be adjusted according to the type of ARMA coefficients. For instance, when both $a^l_{j,p}$ and $b^l_{j,q}$ are real-valued, activations such as leaky ReLU or Tanh can be used. For complex-valued $a^l_{j,p}$ and $b^l_{j,q}$, each parameter is predicted using two separate Conv1D layers, with Tanh and leaky ReLU used to output the phase and magnitude, respectively, which are then combined to form complex $a^l_{j,p}$ and $b^l_{j,q}$. }
		\label{QHM_GAN_ARMA_structure}
		\vspace{-0.3cm}
	\end{figure*}

	As described in the previous subsection, the DNN is utilized to generate the ARMA coefficients ($a_p$ and $b_q$). We evenly divide them into $r$ parts, i.e., $a_{j,p}$ and $b_{j,q}$ ($j=1,\cdots,r$). The frequency response in Eq. (\ref{eq15}) can be interpreted as a cascade of mini-ARMA models, which can be rewritten as 
	\begin{align}\label{eq22}
	\resizebox{0.892\hsize}{!}{$\begin{aligned}
		H(t_l,\omega) = G_l\prod_{j=1}^{r}\tilde{H}_j(t_l,\omega)=G_l\prod_{j=1}^{r}\frac{1+\sum_{q=1}^{Q/r}b^l_{j,q}e^{-i\omega q}}{1+\sum_{p=1}^{P/r}a^l_{j,p}e^{-i\omega p}}, 
		\end{aligned}$}
	\end{align}
	where $\tilde{H}_j(t_l,\omega)$ is the mini-ARMA model. Each $\tilde{H}_j(t_l,\omega)$ represents a stage in the cascade, contributing specific spectra to the overall model. By cascading these mini-models sequentially, $	H(t_l,\omega)$ becomes suitable for capturing the complex structures of speech, while $G_l$ represents the time-varying amplitude gain.
	Accordingly, the amplitude and phase delay of the $k$-th component at the $l$-th frame can be determined by 
	\begin{equation}\label{eq23}
	\resizebox{0.892\hsize}{!}{$\begin{aligned}
		\hat{A}_k(t_l) = G_l\prod_{j=1}^{r}|\tilde{H}_j(t_l,\omega_k)|, \angle {H}(t_l,\omega_k) = \sum_{j=1}^{r}\angle \tilde{H}_j(t_l,\omega_k).
		\end{aligned}$}
	\end{equation}
	Subsequently, adopting the interpolation algorithms similar to those of QHM-GAN, the instantaneous amplitude and phase can be obtained for speech waveform synthesis. We refer to such an improved QHM-GAN as QHARMA-GAN. Note that, since QHARMA-GAN models the time-varying spectral envelopes, $f_0$ need not be fed into the DNN, and the FCNN is no longer required. The framework of QHARMA-GAN is illustrated in Fig. \ref{QHM_GAN_ARMA_structure} while the details of the synthesis process of QHARMA-GAN are elaborated in Algorithm \ref{alg1_1}.
	
	In terms of training, our preliminary experiment shows that the multi-resolution discriminator (MRD) and multi-period discriminator (MPD) are primarily effective in guiding the generator to learn voiced speech (harmonic), while the multi-scale discriminator (MSD) excels in guiding the generator to produce high-quality unvoiced speech (stochastic). Thus, we incorporate these three discriminators to build the discriminator for QHARMA-GAN. Moreover, a large frame-shift will degrade the synthesis quality owing to the increase in the amount of the missing of the information between frames. To address this issue, a transpose convolution can be applied to upsample the ARMA coefficients. Then, using an upsampled $f_0$ enables a more meticulous speech waveform resynthesis.
	\begin{algorithm}[!t]
		\caption{Speech Synthesis based on QHARMA-GAN.}
		\begin{algorithmic}
			\STATE {\textbf{Step 1: Preprocessing and Setting}\\
				Extract the $\hat{f}_k$ and mel-spectrogram from speech $x(t)$;
				set sampling rate $f_s$, harmonic number $K$ and frame-shift $\Delta t$; 
				obtain $a^l_{j,p}$ and $b^l_{j,q}$ from DNN;}
			
			\STATE {\textbf{Step 2: Instantaneous amplitude and phase}\\
				Get framewise amplitude and phase delay by Eq. (\ref{eq22})-(\ref{eq23});\\
				Compute unwrapped framewise phase $\hat{\varphi}_k(t_l)$ by Eq. (\ref{eq17});\\
				Cubically interpolate $\hat{\varphi}_k(t_l)$ into $\hat{\varphi}_k(t)$;\\
				Linearly interpolate $\hat{A}_k(t_l)$ into $\hat{A}_k(t)$;\\
			}
			\STATE {\textbf{Step 3: Generation}\\
				\STATE $\hat{x}(t) \leftarrow \sum_{k=-K}^{K}\hat{A}_k(t)e^{i\hat{\varphi}_k(t)}$;
			}
			\STATE {\textbf{Output:} $\hat{x}(t)$}
		\end{algorithmic}
	
		\label{alg1_1}
	\end{algorithm}

\begin{figure}[!t]\centering
		\includegraphics[width=8.4cm]{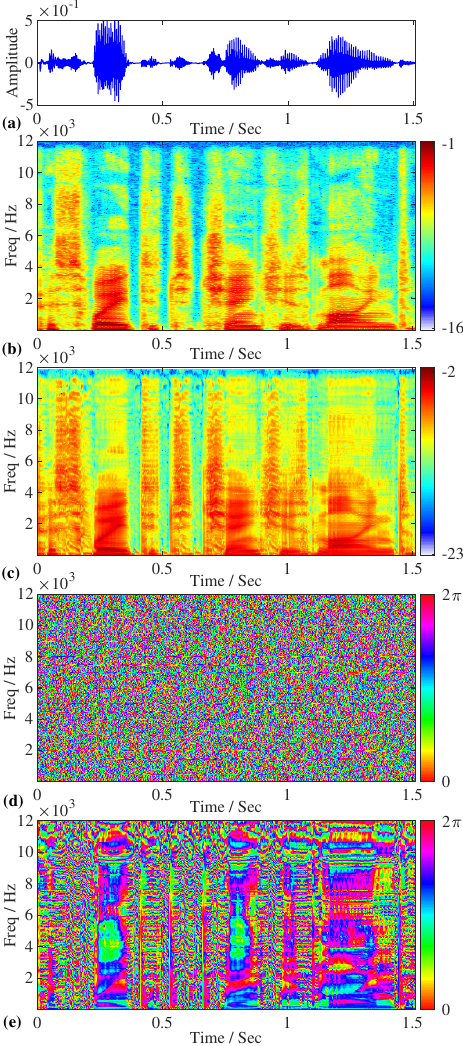}
		\vspace{-0.3cm}
		\caption{(a) Ground truth of utterance sample. Magnitude spectra of (b) ground truth and (c) corresponding ARMA response. Phase spectra of (d) ground truth and (e) corresponding ARMA response.}
		\label{spectral}
		\vspace{-0.3cm}
	\end{figure}

	To show the effectiveness of QHARMA-GAN in resonance characteristic modeling, we illustrate the ground truth of an utterance sample in Fig. \ref{spectral}(a). Furthermore, the decibel-type magnitude, and wrapped phase spectra of the speech signal and corresponding ARMA model response are shown in Fig. \ref{spectral}(b)--(e), where the orders $P = 128$ and $Q = 128$ with $r=8$. Compared to the magnitude and phase of the ground truth in Figs. \ref{spectral}(b) and (d), Figs. \ref{spectral}(c) and (e) show a smooth complex spectral envelope, which can smoothly depict the amplitude and phase of harmonics with the arbitrary $f_0$, demonstrating that QHARMA-GAN can extract the resonance characteristics to remove the excitation signal.

	\subsection{Speech Modification based on QHARMA-GAN}	
	QHARMA-GAN models the speech (including unvoiced and voiced parts) as a combination of quasi-harmonics. Thus, it is straightforward to modify the quasi-harmonics to achieve speech modification \cite{di1998waveform}, \cite{kafentzis2014adaptive}, i.e., modify the frequency for pitch shifting and the duration time for time stretching. Since the spectral envelope is modeled and determined, the harmonics, whose frequencies are multiples of the given $f_0$ below the Nyquist frequency ($f_k = k f_0 < \frac{f_s}{2}$), will be generated for speech synthesis. Thus, increasing $f_0$ is bound to reduce the number of harmonics, inevitably degrading the quality of unvoiced speech, because unvoiced speech, as stochastic signals, requires a sufficient number of frequency components to be well fitted. To overcome this limitation, we are inspired by the harmonic plus noise model \cite{stylianou1996harmonic},\cite{pantazis2008improving} to employ the voiced/unvoiced (V/UV) mask to selectively modify voiced speech for pitch- and time-scale modifications, the specific algorithms of which are given below.

	According to Eq. (\ref{eq9}), the amplitudes and phases of quasi-harmonics are essential to speech modification. On basis of the properties of ARMA, we can control $G_l$ to adjust the loudness and modify both amplitude and phase by inputting a modified $f_0$ into ARMA.
	Specifically, we only modify the voiced part while keeping the unvoiced part fixed, ensuring the quality of unvoiced speech. To ensure the power of speech fixed during modification, we can control the gain according to the number of harmonics of the modified speech. Denoting the time-scale factor as $\beta_l$ and the pitch-scale factor as $\rho_l$ at the $l$-th frame, we specify the modification in the following part:
	\subsubsection{Frequency scaling and amplitude estimation} First, for time scaling, the index for interpolation (denoted as $\hat{t}_l$) should be modified by the time-scale factor $\beta$ as
	\begin{align}
	\hat{t}_l = \hat{t}_0 + \sum_{i=1}^l \beta_i (t_i - t_{i-1}), \ \ \hat{t}_0=0,
	\label{eq24}
	\end{align}
	For pitch scaling, we shift frequencies by the pitch-scale factor $\rho$ as
	\begin{align}
	\text{Voiced: }\hat{f}_{k,v}^l = \rho_l \hat{f}_k^l, \ \ \text{Unvoiced: } \hat{f}_{k,uv}^l = \hat{f}_k^l.
	\label{eq25}
	\end{align}
	Then, their amplitude can be obtained by Eqs. (\ref{eq22})-(\ref{eq23}) as
	\begin{subequations}
		\begin{align}\label{eq26a}
		\hat{A}_k^v(\hat{t}_l) &= VUV_l \times G\prod_{j=1}^{r}|\tilde{H}_j(t_l,2\pi \hat{f}_{k,v}^l)|,\\
		\hat{A}_k^{uv}(\hat{t}_l) &=(1-VUV_l) \times G\prod_{j=1}^{r}|\tilde{H}_j(t_l, 2\pi \hat{f}_{k,uv}^l)|,\label{eq26b}
		\end{align}
		\label{eq26}
	\end{subequations}
    where $VUV_l$ is a binary coefficient. When the speech of the $l$-th frame is voiced, $VUV_l=1$, otherwise, $VUV_l=0$. 
	The linear interpolation is employed to obtain the instantaneous versions of $\hat{A}_k^v(\hat{t}_l)$ and $\hat{A}_k^{uv}(\hat{t}_l)$, i.e., $\hat{A}_k^v(\hat{t})$ and $\hat{A}_k^{uv}(\hat{t})$. 
	
	\subsubsection{Phase estimation}
	To ensure the smoothness of the phase, we separately calculate the instantaneous phase of voiced speech and unvoiced speech. We compute the phase delay by substituting the frequencies in Eq. (\ref{eq25}) to Eq. (\ref{eq23}). The phase delay will be used to obtain the framewise phase for voiced speech and unvoiced speech by Eq. (\ref{eq17}), respectively, i.e., $\varphi^u_{k}(\hat{t}_l)$ and $\varphi^{uv}_{k}(\hat{t}_l)$. Note that, the phase of the excitation signal in Eq. (\ref{eq18}) can be quickly obtained by Eq. (\ref{eq11}) as 
	\begin{align}
	\varphi^u_{k}(\hat{t}_l) &= \int_{0}^{t_{l}}{2\pi {\hat{f}_{k}}(u)} \text{d}u \notag \\
	&\approx \pi\sum_{i=1}^{l}{[\hat{f}_k^{i-1} + \hat{f}_k^{i}] (t_{i} - t_{i-1})\beta_l},
	\label{eq27}
	\end{align}
	where $\hat{f}_k$ should be $\hat{f}_{k,v}^l$ and $\hat{f}_{k,uv}^l$. After obtaining the framewise phase, the cubic interpolation is employed to compute the instantaneous phase of voiced speech and unvoiced speech, i.e., $\hat{\varphi}_{k,v}(\hat{t})$ and $\hat{\varphi}_{k,uv}(\hat{t})$. 
	
	\subsubsection{Synthesis}
	The unvoiced speech and voiced speech can be separately synthesized and formed into the final speech by 
	\begin{align}
	\hat{x}(\hat{t}) =& \hat{x}_{uv}(\hat{t}) + \hat{x}_v(\hat{t}) \notag \\
	=& \sum_{k=-K}^{K}\hat{A}_k^v(\hat{t})e^{i\hat{\varphi}_{k,v}(\hat{t})} + \sum_{k=-K}^{K}\hat{A}_k^{uv}(\hat{t})e^{i\hat{\varphi}_{k,uv}(\hat{t})}.
	\label{eq28}
	\end{align}
	To illustrate the speech modification more intuitively, we present the specific process as Algorithm \ref{alg2}.

	\begin{algorithm}
		\caption{Time- and pitch-scale modification.}
		\begin{algorithmic}
			\STATE {\textbf{Step 1: Preprocessing and Setting}\\
				Train the QHARMA-GAN with dataset and infer the ARMA coefficients from DNN part; set time-scale factor $\beta^l$ and pitch-scale factor $\rho^l$;
			}
			\STATE {\textbf{Step 2: Modifying Frequency}\\
				Scale the index for time scaling by Eq. (\ref{eq24}); \\
				$\hat{f}_{k,v}^l \leftarrow \rho_l \hat{f}_k^l, \ \  \hat{f}_{k,uv}^l \leftarrow \hat{f}_k^l$;
				
			}
			\STATE {\textbf{Step 3: Modified Amplitude Estimation}\\
				Obtain the modified amplitudes of voiced speech and unvoiced speech $\hat{A}_k^v(\hat{t}_l)$ and $\hat{A}_k^{uv}(\hat{t}_l)$ by Eq. (\ref{eq26});\\
				Apply linear interpolation to get $\hat{A}_k^v(\hat{t})$ and $\hat{A}_k^{uv}(\hat{t})$;
			}
			\STATE {\textbf{Step 4: Phase Estimation}\\
				Substitute $\hat{f}_{k,v}^l$ and $\hat{f}_{k,uv}^l$ into Eqs. (\ref{eq22})-(\ref{eq23}) for phase delay;\\
				Use Eq. (\ref{eq27}) to get the phase of excitation signal;\\
				Compute modified unwrapped framewise phase by Eq. (\ref{eq17});\\
				Apply cubic interpolation to get $\hat{\varphi}_{k,v}(\hat{t})$ and $\hat{\varphi}_{k,uv}(\hat{t})$;
			}
			\STATE {\textbf{Step 5: Synthesis}\\
				Synthesize the speech by Eq. (\ref{eq28})
			}
			
			\STATE {\textbf{Output:} $\tilde{x}(t')$}
		\end{algorithmic}
		\label{alg2}
	\end{algorithm}

		\begin{figure}[!b]\centering
		\includegraphics[width=8.4cm]{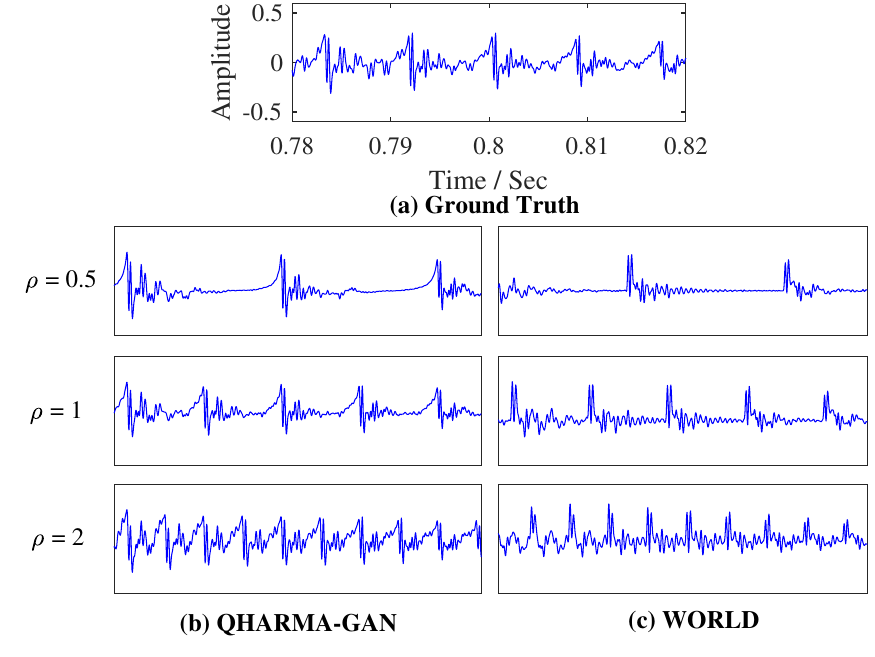}
		\vspace{-0.3cm}
		\caption{(a) The waveform of ground truth in Fig. \ref{spectral} at $t= 0.8$ sec. Denoting pitch scale factor as $\rho$, the waveforms of speech signals modified by (b) QHARMA-GAN and (c) WORLD with different $\rho$s.}
		\label{base_sig}
		\vspace{-0.3cm}
	\end{figure}

	To show the performance of QHARMA-GAN in speech modification, we employ QHARMA-GAN and WORLD to modify the signal in Fig. \ref{spectral}(a), where the enlarged views of synthesized or modified waveforms at $t=0.8$ sec are depicted in Fig. \ref{base_sig}. Apparently, compared with WORLD, the waveform shape of the synthesized or modified speech signals aligns as closely as possible with the ground truth since QHARMA-GAN is capable of recovering the phase information of the speech signals while WORLD lacks phase modeling and synthesizes the waveform only using the magnitude of the spectral envelope. Even during pitch-scale modification, the shape of the modified speech waveform obtained by QHARMA-GAN remains highly consistent with the ground truth. In contrast, WORLD synthesizes waveforms directly from the magnitude spectrum, ignoring phase information, which leads to a significant deviation of waveform shape from the ground truth during both speech synthesis and modification. Thus, QHARMA-GAN's ability to preserve waveform shape during signal generation and modification is undoubtedly a significant advantage for improving speech synthesis quality.

	\section{Experimental Study}
	Several experiments have been conducted to evaluate the performances of various well-known vocoders and our proposed methods in terms of synthesis quality, generation speed, etc.. The results are quantized and compared in this section. 
	\subsection{Experimental Conditions}
	To validate the generalization ability and the performance in speaker-independent synthesis, we separately trained the models using 
	two multi-speaker corpora, i.e., the VCTK corpus \cite{yamagishi2019cstr} composed of 110 speakers and the Japanese Versatile Speech (JVS) corpus \cite{takamichi2019jvs} composed of 100 speakers. Both datasets were collected from male and female speakers, with a sampling rate set to 24000 Hz.

	In terms of the model candidates, we selected HiFi-GAN, Vocos \cite{Vocos}, and hn-NSF \cite{wang2019neural} as representatives of neural vocoders, and QHM and WORLD as representatives of CSP vocoders, to compare with our proposed QHM-GAN and QHARMA-GAN. Additionally, to improve computational efficiency, we simplified the DNN part of QHARMA-GAN and named it QHARMA-GAN-small, whose results were also included in the comparison. The overviews of these models are as follows 
	\begin{itemize} 
		\item [ 1)] \textbf{WORLD}: A CSP source-filter vocoder with a flexible controllability of acoustic features, such as $f_0$, that extracts acoustic features and resynthesizes the speech with a reasonable quality.
		\item [ 2)] \textbf{QHM}: A quasi-harmonic modeling method with a frequency correction mechanism. QHM extracts the framewise complex amplitudes of each quasi-harmonic to adaptively correct their frequency. Using the complex amplitude and corrected frequency can resynthesize the speech waveform perfectly and modify the speech from the perspectives of time and frequency along with DAP. 
		\item [ 3)] \textbf{HiFi-GAN}: A well-known high-quality neural vocoder, which applies progressive transposed convolutions for upsampling and MRF modules to process the output after each upsampling stage, ultimately producing the speech. During training, MSD and MPD are employed to evaluate the result from the generator and guide the learning process.
		\item [ 4)] \textbf{Vocos}: An efficient neural vocoder, which mainly adapts ConvNeXt \cite{ConvNeXt} to generate complex spectrograms without upsampling and synthesizes speech waveforms with iSTFT. In \cite{yoneyama2024wavehax}, a pitch-controllable Vocos is developed using external harmonic inputs from an $f_0$ prior, enabling $f_0$ editing in a source-filter framework. Inspired by this, we adopt the same design for $f_0$ extrapolation experiments. During training, MRD and MPD are used to evaluate the generator output against the ground truth.
		\item [ 5)] \textbf{Hn-NSF}: A neural source-filter vocoder that inputs a harmonic-plus-noise signal generated from $f_0$ prior, enabling $f_0$ editing. The source signal is passed trough a learnable filter to produce the final waveform. During training, the HiFi-GAN discriminator is adopted to guide the generator towards high-quality speech synthesis.
		\item [ 6)] \textbf{QHM-GAN}: Our previous method with four MRF modules. Each MRF module has three kernels, where each kernel processes with dilation rates of 1, 3, and 5. The DNN part converts a mel-spectrogram to amplitudes and phase compensations coefficients, with which the speech can be resynthesized by CSP part, as shown in Fig. \ref{QHM_GAN_structure}. MRD and MPD are employed during training.
		\item [ 7)] \textbf{QHARMA-GAN}: Our proposed method with four MRF modules. The configuration of MRF is the same as that of QHM-GAN. QHARMA-GAN uses a mel-spectrogram to generate ARMA coefficients for modeling the spectral envelope, with which the speech can be resynthesized and arbitrarily modified with the pre-extracted individual frequency. During training, MRD, MSD, and MPD are employed to improve the training.
		\item [ 8)] \textbf{QHARMA-GAN-small}: The simplified version of QHARMA-GAN with only three MRF modules. Each MRF module has three kernels, where each kernel processes only with the dilation rate of 1. MRD, MSD, and MPD are employed during training.
		
	\end{itemize} 
	
	With regard to the metrics for evaluating the performances of the various methods, we utilize the following metrics to quantize the performances. Note that the upward arrow indicates that a higher value for the metric signifies better performance, while the downward arrow means that a lower value signifies better performance.
	\begin{itemize} 
		
		\item [ 1)] \textbf{V/UV rate} [\%] $ \downarrow$: This metric quantifies the proportion of incorrect detection of speech segments as either voiced speech (produced with vocal cord vibration) or unvoiced speech (produced with airflow). V/UV detections were determined on the basis of whether $f_0$ detected by Harvest \cite{morise2017harvest} is equal to zero; values of zero indicate unvoiced speech, while nonzero values indicate voiced speech. The V/UV detections of generated speech and ground truth will be compared to compute the V/UV rate.
		\item [ 2)] \textbf{$f_0$ RMSE} [Hz] $ \downarrow$: This metric is the root mean squared
		error (RMSE) of logarithmic $f_0$s between generated speech and ground truth detected by Harvest. It is used to assess the reconstruction and modification of pitch. Note that, for the measurement of speech modification, the RMSE will be determined between $f_0$ modified speech and scaled $f_0$ of ground truth, as
		\begin{align}\label{eq29}
		\resizebox{0.80\hsize}{!}{$\begin{aligned}
		RMSE = \sqrt{\frac{1}{L_v}\sum\limits_{l=1}^{L_v}{\left[\log (f^{gen}_{0,l})-\log (\rho_l f^{ref}_{0,l}) \right]^{2}}},
		\end{aligned}$}
		\end{align}
		where $ f^{gen}_{0,l}$ and $ f^{ref}_{0,l}$ are respectively the detected $f_0$ values of synthesized speech and ground truth at the $l$-th voiced frame ($l = 1,...,L_v$, $L_v$ is the number of voiced frames), and the pitch-scale factor is equal to 1 in the case of synthesis, i.e., $\rho_l  \equiv 1$.
		\item [ 3)] \textbf{PESQ} $ \uparrow$: This metric is used to assess speech quality by measuring the spectral distance between synthesized speech and the reference, offering a correlation with human auditory perception.
		\item [ 4)] \textbf{UTMOS}\footnote{https://github.com/sarulab-speech/UTMOS22} $ \uparrow$: This metric provides the estimate of subjective speech quality, i.e., mean opinion scores (MOS), with a pretrained neural network \cite{saeki2022utmos}. It serves for overall speech quality assessment to mainly screen the best baseline.
			
		\item [ 5)] \textbf{MCD} $\downarrow$: This metric measures the distance between mel-cepstral coefficients of the generated and reference waveforms, computed as
		\begin{align}
		MCD(v_{\text{gen}},v_{\text{ref}})= \frac{10\sqrt{2}}{\ln{10}}\sqrt{\sum_{d=1}^{24}(v_{\text{gen}}^d-v_{\text{ref}}^d)^2},
		\label{MCD}
		\end{align}
		 where $v_{\text{gen}}$ and $v_{\text{ref}}$ are the mel cepstrum coefficients of synthesized and reference speech signals and $d$ is the dimension index. It provides an objective measure of spectral envelope distortion and is commonly used to evaluate the fidelity of synthesized speech.
	 
		\item [ 6)] \textbf{MOS} $ \uparrow$: This indicator assesses subjective speech quality based on evaluations from diverse human listeners. Taking into account factors such as sound quality, clarity, and intelligibility, participants rated the overall naturalness of the speech on a scale of 1 (poor) to 5 (excellent).
		\item [ 7)] \textbf{RTF} $ \downarrow$: This metric is called the real-time factor (RTF) and demonstrates the processing speed in speech analysis and generation, indicating whether the method can be applied in real-time processing.
		
	\end{itemize}

	All scores were averaged by evaluating the results across multiple sets of speech samples. Notably, the results from the two datasets were assessed independently. Objective evaluations were averaged from measurements of the 1885 utterances from VCTK and 980 utterances from JVS, while each subjective MOS value was averaged over evaluations by 20 recruited participants, who rated 12 randomly selected utterances, a total of 72 samples, i.e., $12\times6$ methods, for each dataset. Audio samples in the following experiments and code are available from our demo site\footnote{https://chen-shaowen.github.io/QHARMA-GAN-Demo/}.

	\begin{table}[!t]\centering
	\caption{Results of objective and subjective evaluations for speech synthesis. The UTMOS values of the ground truth samples for VCTK and JVS were 4.04 and 3.63, respectively.}
	\setlength\tabcolsep{1pt} 
	\centering
	\label{table_pre1}
	\renewcommand{\arraystretch}{1.3} 
	\begin{tabular}{l|c|c|c|c|c|c}
		\toprule 
		Metric & Dataset & HiFi-GAN & Vocos & \makecell{hn-NSF} &  \makecell{ QHM\\-GAN} & \makecell{QHARMA\\-GAN} \\
		\midrule
		\multirow{2}{*}{V/UV rate {[}\%{]} $\downarrow$}  & VCTK & \textbf{11}&\textbf{11}&\textbf{11} &13&\textbf{11}    \\
		& JVS  & 11&\textbf{9}& \textbf{9} &14&\textbf{9}      \\ \hline
		\multirow{2}{*}{$f_0$ RMSE {[}Hz{]} $\downarrow$ }& VCTK & \textbf{0.06}& \textbf{0.06} & \textbf{0.06} &\textbf{0.06}&\textbf{0.06}    \\ 
		& JVS  &  0.11 & 0.11 & 0.10 & \textbf{0.09} & \textbf{0.09}    \\\hline
		\multirow{2}{*}{PESQ $\uparrow$}	      & VCTK & 3.14 & \textbf{3.15} & 2.91 & 3.00 & 3.14      \\
		& JVS  & 3.29 & \textbf{3.42} & 3.23& 3.29 & 3.26     \\ \hline
		\multirow{2}{*}{MCD $\downarrow$}	      & VCTK & \textbf{3.61}& 3.62 & 4.21 &4.14&4.09     \\
		& JVS  & 3.6 & \textbf{3.45} & 3.91 &4.08 & 4.00     \\ \hline
		\multirow{2}{*}{UTMOS $\uparrow$}         & VCTK &\textbf{3.91}& 3.89 & 3.76 &3.50&3.76     \\
		& JVS  & 3.16& 3.25 & 3.21 & 3.15 & \textbf{3.30} \\ 
		\bottomrule
	\end{tabular}
\vspace{-0.6cm}
\end{table}

	\subsection{Evaluation of Speech Synthesis and Modification}
	
	In this part, the performances of all methods are compared in terms of synthesis and pitch-scale modification. We adopt a two-stage evaluation strategy. In the first stage, we use only objective indicators to screen out the best-performing baseline (e.g., HiFi-GAN, Vocos or hn-NSF) for other neural models and our best model (e.g., QHM-GAN, QHARMA-GAN). In the second stage, these selected models are compared in detail with conventional methods such as WORLD and QHM.

	\begin{table}[!t]\centering
		\caption{Results of objective evaluations for $f_0$ modification. The UTMOS of the ground truth samples without modification was the same as that shown in Table \ref{table_pre1}.}
		\setlength\tabcolsep{2pt} 
		\vspace{-0.1cm}
		\centering
		\label{table_pre2}
		\renewcommand{\arraystretch}{1.3} 
		\begin{tabular}{l|c|l|c|c|c|c}
			\toprule 
			Metric & Dataset & Scale & Vocos & \makecell{hn-NSF} &   \makecell{QHM\\-GAN}& \makecell{QHARMA\\-GAN} \\
			\midrule
			\multirow{8}{*}{\makecell[l]{V/UV rate {[}\%{]} $\downarrow$}} & \multirow{4}{*}{VCTK} & $\rho_l \equiv 2^{-1}$   & 15& 15 & 18&\textbf{14}  \\
			&						 & $\rho_l \equiv 2^{-0.5}$ &\textbf{12} & \textbf{12} &16&13  \\
			&						 & $\rho_l \equiv 2^{0.5}$  &14 & 14 &17&\textbf{11}  \\
			&						 & $\rho_l \equiv 2^{1}$     &\textbf{13} & 27 &18&\textbf{13}  \\
			\cmidrule(lr){2-7}
			& \multirow{4}{*}{JVS}  & $\rho_l \equiv 2^{-1}$     & 15 & 19 & 17 & \textbf{14}  \\
			&						 & $\rho_l \equiv 2^{-0.5}$  & 14 & 14 & 13 & \textbf{12}  \\
			&						 & $\rho_l \equiv 2^{0.5}$    & 13 & 13 & 13 & \textbf{12}  \\
			&						 & $\rho_l \equiv 2^{1}$      & \textbf{14} & 15 & 15 & \textbf{14}  \\ \hline
			\multirow{8}{*}{\makecell[l]{$f_0$ RMSE  {[}Hz{]} $\downarrow$}}&\multirow{4}{*}{VCTK}&$\rho_l \equiv 2^{-1}$  & 0.65 & 0.29 &0.32&\textbf{0.11}  \\
			&					   &$\rho_l \equiv 2^{-0.5}$& 0.33 & 0.09 &\textbf{0.08}&\textbf{0.08}  \\
			&					   &$\rho_l \equiv 2^{0.5}$  & 0.34 & 0.10 &\textbf{0.08}&0.09  \\
			&					   &$\rho_l \equiv 2^{1}$    & 0.67 & 0.32 &\textbf{0.07}&0.10  \\
			\cmidrule(lr){2-7}
			&\multirow{4}{*}{JVS} & $\rho_l \equiv 2^{-1}$     & 0.64 & 0.28 & 0.20& \textbf{0.15}  \\
			&					   & $\rho_l \equiv 2^{-0.5}$    & 0.34 & 0.15 & 0.14 & \textbf{0.12}  \\
			&					   & $\rho_l \equiv 2^{0.5}$   & 0.35 & 0.15 & \textbf{0.14} & \textbf{0.14}  \\
			&					   & $\rho_l \equiv 2^{1}$      & 0.69 & 0.19 & \textbf{0.11} & 0.12  \\ \hline
			\multirow{8}{*}{ \makecell[l]{MCD {[dB]} $\downarrow$}}&\multirow{4}{*}{VCTK}&$\rho_l \equiv 2^{-1}$   & \textbf{3.59} & 5.52 &8.81&5.28 \\
			&					   &$\rho_l \equiv 2^{-0.5}$ & \textbf{3.76} & 4.60 &5.51&4.31  \\
			&					   &$\rho_l \equiv 2^{0.5}$ & \textbf{3.78} & 4.98 &4.96&4.67 \\
			&					   &$\rho_l \equiv 2^{1}$    & \textbf{3.95} & 5.83 &7.27&5.21 \\
			\cmidrule(lr){2-7}
			&\multirow{4}{*}{JVS} & $\rho_l \equiv 2^{-1}$     & \textbf{3.62} & 5.19 & 6.97 & 4.91  \\
			&					   & $\rho_l \equiv 2^{-0.5}$  & \textbf{3.63} & 4.49 & 5.51 & 4.29  \\
			&					   & $\rho_l \equiv 2^{0.5}$    & \textbf{3.82} & 4.78 & 5.34 & 4.72  \\
			&					   & $\rho_l \equiv 2^{1}$      & \textbf{3.88} & 5.73 & 6.71 & 5.54  \\ \hline
			\multirow{8}{*}{UTMOS $\uparrow$}&\multirow{4}{*}{VCTK}&$\rho_l \equiv 2^{-1}$   & \textbf{3.49} & 3.16 &1.53&3.14  \\
			&					   &$\rho_l \equiv 2^{-0.5}$ & 2.78 & 3.60 &2.22&\textbf{3.61}  \\
			&					   &$\rho_l \equiv 2^{0.5}$  & 2.81 & 2.93 &2.47&\textbf{3.25}  \\
			&					   &$\rho_l \equiv 2^{1}$    & \textbf{3.03} & 2.00 &1.62&2.68  \\
			\cmidrule(lr){2-7}
			&\multirow{4}{*}{JVS} & $\rho_l \equiv 2^{-1}$    & \textbf{2.80} & 2.34 & 1.32 & 2.66  \\
			&					   & $\rho_l \equiv 2^{-0.5}$  & 1.83 & 2.79 & 1.57 & \textbf{3.05}  \\
			&					   & $\rho_l \equiv 2^{0.5}$   & 1.73 & 2.22 & 1.87 & \textbf{2.23} \\
			&					   & $\rho_l \equiv 2^{1}$      & 1.83 & 1.37 & 1.46 & \textbf{1.84}  \\ 
			\bottomrule
		\end{tabular}
	\vspace{-0.6cm}
	\end{table}

	First, a preliminary experiment is conducted, where the objective evaluation results on the VCTK and JVS datasets are presented in Tables \ref{table_pre1} and \ref{table_pre2}. Table \ref{table_pre1} assesses the synthesis performances of all neural vocoders, showing that HiFi-GAN and Vocos exhibit comparable performance, with slight trade-offs across metrics and datasets. Hn-NSF lags behind Vocos and HiFi-GAN in both MCD and UTMOS, indicating relatively lower quality in spectral and perceptual aspects. Moreover, among the proposed models, QHARMA-GAN consistently generates higher-quality speech, notably achieving the highest UTMOS on the JVS dataset. Therefore, considering its widespread adoptions in recent studies as a standard baseline and its performance comparable to that of Vocos, HiFi-GAN is chosen as the baseline to compare with QHARMA-GAN in the main synthesis experiment.

	Table \ref{table_pre2} evaluates the $f_0$ extrapolation performances under various pitch-scale factors: $\rho_l\equiv2^{-1}$, $\rho_l\equiv2^{-0.5}$, $\rho_l\equiv2^{0.5}$, and $\rho_l\equiv2^{1}$. Vocos achieves the best MCD, while its UTMOS is comparable to that of hn-NSF. However, Vocos exhibits significantly high $f_0$ RMSE, suggesting its failure in pitch modification\footnote{The audio samples are available in the demo site: https://chen-shaowen.github.io/QHARMA-GAN-Demo/.}. In contrast, the relatively stable $f_0$ RMSE of hn-NSF indicates its ability to modify pitch. Nevertheless, it still struggles to maintain high quality, particularly under extreme pitch-scale factors (i.e., when $\rho \equiv 2^{-1}$ or $2^{1}$). Focusing on the proposed methods, QHM-GAN achieves satisfactory $f_0$ RMSE, indicating effective pitch control. However, it yields the worst MCD values, as it fails to model resonance characteristics and accurately predict the amplitude of each component. This leads to notable spectral distortions and degraded perceptual quality, as reflected by its lowest UTMOS score. In contrast, the excellent V/UV rates and $f_0$ RMSEs of QHARMA-GAN imply its ability to maintain accurate $f_0$ extrapolation while minimizing spectral distortion across all pitch-scale factors. Its higher UTMOS further confirms that QHARMA-GAN possesses the most effective $f_0$ extrapolation among the compared methods.
	
	\begin{table*}[!b]\centering
		\caption{Results of objective and subjective evaluations. The MOS values of the ground truth samples for VCTK and JVS datasets were $4.27\pm0.022$ and $3.88\pm0.038$, respectively.}
		\setlength\tabcolsep{4pt}
		\vspace{-0.2cm} 
		\centering
		\label{table_1}
		\renewcommand{\arraystretch}{1.3} 
		\begin{tabular}{l|c|c|c|c|c|c}
			\toprule 
			Metric & Dataset & WORLD & QHM & HiFi-GAN & QHARMA-GAN & QHARMA-GAN-small \\
			\midrule
			\multirow{2}{*}{V/UV rate {[}\%{]} $\downarrow$}  & VCTK & \textbf{11} & 13 & \textbf{11} & \textbf{11} & 12    \\
			& JVS  & 12 & 15 & 11 & \textbf{9} & 10      \\ \hline
			\multirow{2}{*}{$f_0$ RMSE {[}Hz{]} $\downarrow$ }& VCTK & 0.06 & \textbf{0.03} & 0.06 & 0.06 & 0.06      \\ 
			& JVS  & 0.11 & \textbf{0.05} & 0.11 & 0.09 & 0.09      \\\hline
			\multirow{2}{*}{PESQ $\uparrow$}	      & VCTK & 2.54 & \textbf{3.45} & 3.14 & 2.72 & 2.49      \\
			& JVS  & 2.97 & \textbf{3.61} & 3.29 & 3.26 & 3.18      \\ \hline
			\multirow{2}{*}{MOS $\uparrow$}           & VCTK & $4.12\pm0.025$ & $4.07\pm0.035$ & $4.08\pm0.028$ & $\bm{ 4.21\pm0.025 }$ & $4.10\pm0.054$    \\
			& JVS  & $3.67\pm0.042$ & $3.40\pm0.047$ & $3.64\pm0.030$ & $\bm{3.80\pm0.029}$ & $3.68\pm0.030$           \\
			\bottomrule
		\end{tabular}
	\vspace{-0.6cm}
	\end{table*}
	
\begin{figure}[!t]\centering
		\vspace{-0.6cm}
		\includegraphics[width=8.9cm]{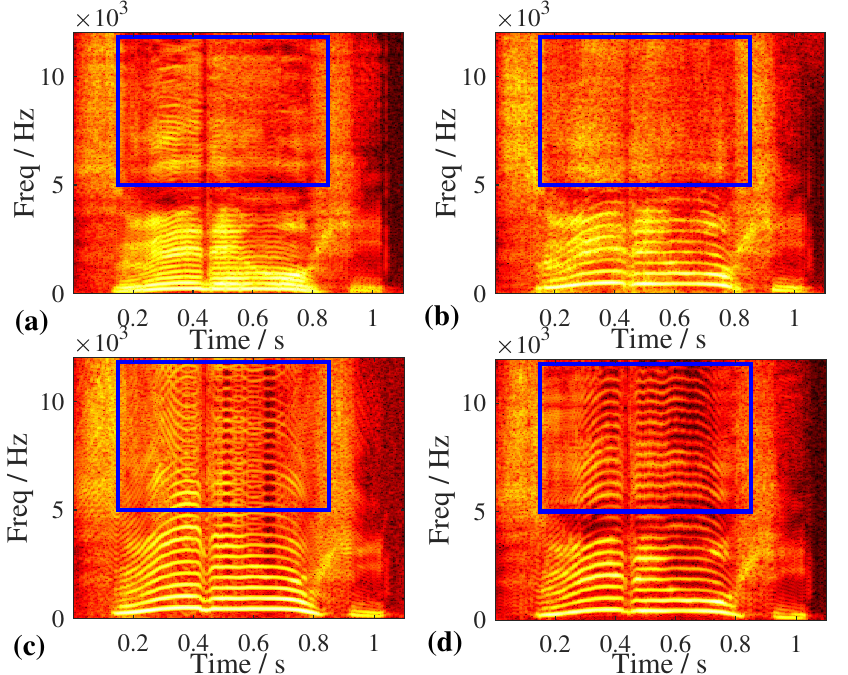}
		\caption{The spectrograms of the speech generated by (a) Vocos, (b) hn-NSF, (c) QHM-GAN, and (d) QHARMA-GAN with $\rho \equiv 2^{1}$.}
		\label{specs}
		\vspace{-0.4cm}
	\end{figure}

	To further illustrate the performance of $f_0$ extrapolation, the spectrograms of a pitch-scaled speech sample ($\rho \equiv 2^{1}$) by various methods are presented in Fig. \ref{specs}. Clearly, Vocos fails to generate the speech with correct pitch. Moreover, both Vocos and hn-NSF struggles to reconstruct the high-frequency harmonics, whereas QHM-GAN and QHARMA-GAN can synthesize such components harmonically, resulting in sharp high-frequencies trajectories in the spectrogram. However, the QHM-GAN seems to exhibit aliasing pattern. Thus, QHARMA-GAN achieve the best overall performance in $f_0$ extrapolation.
	Based on the above analysis, we only compare QHARMA-GAN with conventional vocoders (QHM and WORLD) in terms of $f_0$ extrapolation in the main experiment.

	Table \ref{table_1} lists the evaluation results in main experiments, including objective and subjective metrics. In terms of objective results, it is apparent that QHM outperforms other methods, showing the highest PESQ value and lowest $f_0$ RMSE, on account of directly modeling the speech waveform and adaptively correcting the frequency of each component. Naturally, QHM is able to accurately extract the complex amplitude and perfectly reconstruct the speech. However, QHM has a high V/UV rate, since QHM has a limited ability to correct the frequency.
	When the frequency mismatch is significant, it cannot correct the frequency in a single step and requires multiple iterations for a full correction \cite{pantazis2010iterative}. Particularly in the case of analyzing unvoiced speech, the corrected frequencies still exhibit a harmonic-like pattern, making them prone to being misidentified as voiced speech by the pitch detector.
	In contrast, HiFi-GAN shows a slightly lower performance than QHM in terms of PESQ. Although QHARMA-GAN achieves the lowest V/UV rate, it obtains intermediate results in both speech quality and $f_0$ RMSE. 
	
	In terms of subjective MOS results, QHARMA-GAN achieves the highest scores, even higher than the scores of HiFi-GAN and QHM. Since QHM struggles to adjust the frequencies for unvoiced speech, thus, the harmonic-like pattern undoubtedly diminishes the quality of unvoiced speech. On the other hand, HiFi-GAN generates waveforms directly in the time domain, which does not ensure the stability and smoothness of frequencies. As a result, some speech samples generated by HiFi-GAN exhibit noticeable frequency distortions. On the contrary, the synthesis process of QHARMA-GAN ensures frequency stability and smoothness in the resynthesized speech, significantly reducing frequency distortion. Moreover, it can adaptively adjust the phase delay in unvoiced parts based on the input mel-spectrogram, ensuring that these parts exhibit a quasi-harmonic state, thereby improving the perception quality of unvoiced speech. Even QHARMA-GAN-small, whose neural part is simplified, is able to achieve MOS score only slightly lower than that of QHARMA-GAN, implying that such a hybrid framework is superior.

	\begin{table*}[!t]\centering
		\caption{Results of objective and subjective evaluations. The MOS of the ground truth samples without modification was the same as that shown in Table \ref{table_1}.}
		\setlength\tabcolsep{4pt} 
		\vspace{-0.1cm}
		\centering
		\label{table_2}
		\renewcommand{\arraystretch}{1.3} 
		\begin{tabular}{l|c|l|c|c|c|c}
			\toprule 
			Metric & Dataset & Scale & WORLD & QHM & QHARMA-GAN & QHARMA-GAN-small \\
			\midrule
			\multirow{8}{*}{V/UV rate {[}\%{]} $\downarrow$} & \multirow{4}{*}{VCTK} & $\rho_l \equiv 2^{-1}$   &15&18&\textbf{14}&16  \\
			&						 & $\rho_l \equiv 2^{-0.5}$ &14&18&\textbf{13}&14  \\
			&						 & $\rho_l \equiv 2^{0.5}$  &13&19&\textbf{11}&13  \\
			&						 & $\rho_l \equiv 2^{1}$     			&\textbf{13}&21&\textbf{13}&14  \\
			\cmidrule(lr){2-7}
			& \multirow{4}{*}{JVS}  & $\rho_l \equiv 2^{-1}$   & 16 & 16 & \textbf{14} & \textbf{14}  \\
			&						 & $\rho_l \equiv 2^{-0.5}$ & 14 & 14 & \textbf{12} & 13  \\
			&						 & $\rho_l \equiv 2^{0.5}$  & 14 & 18 & 12 & \textbf{11}  \\
			&						 & $\rho_l \equiv 2^{1}$    & 15 & 22 & \textbf{13} & \textbf{13}  \\ \hline
			\multirow{8}{*}{$f_0$ RMSE {[}Hz{]} $\downarrow$}&\multirow{4}{*}{VCTK}&$\rho_l \equiv 2^{-1}$  &0.13& \textbf{0.11} &\textbf{0.11}&\textbf{0.11}  \\
			&					   &$\rho_l \equiv 2^{-0.5}$&0.10& \textbf{0.06} &0.08&0.09  \\
			&					   &$\rho_l \equiv 2^{0.5}$ &0.11& \textbf{0.04} &0.09&0.10  \\
			&					   &$\rho_l \equiv 2^{1}$   &0.16& \textbf{0.04} &0.10&0.11  \\
			\cmidrule(lr){2-7}
			&\multirow{4}{*}{JVS} & $\rho_l \equiv 2^{-1}$   & 0.17 & \textbf{0.13} & 0.15 & 0.17  \\
			&					   & $\rho_l \equiv 2^{-0.5}$ & 0.14 & \textbf{0.08} & 0.12 & 0.13  \\
			&					   & $\rho_l \equiv 2^{0.5}$  & 0.16 & \textbf{0.05} & 0.12 & 0.12  \\
			&					   & $\rho_l \equiv 2^{1}$    & 0.21 & \textbf{0.05} & 0.12 & 0.14  \\ \hline
			\multirow{8}{*}{MOS $\uparrow$}      & \multirow{4}{*}{VCTK} & $\rho_l \equiv 2^{-1}$   & 2.98$\pm$0.056  & 2.43$\pm$0.043 & \textbf{3.01$\pm$0.050} & 2.99$\pm$0.044  \\
			&						 & $\rho_l \equiv 2^{-0.5}$ & 3.84$\pm$0.052  & 3.11$\pm$0.036 & \textbf{3.86$\pm$0.033} & 3.62$\pm$0.057  \\
			&						 & $\rho_l \equiv 2^{0.5}$  & \textbf{3.78$\pm$0.084}  & 3.29$\pm$0.072 & 3.66$\pm$0.066 & 3.29$\pm$0.072  \\
			&						 & $\rho_l \equiv 2^{1}$    & \textbf{2.82$\pm$0.071}  & 2.33$\pm$0.064 & 2.72$\pm$0.057 & 2.69$\pm$0.053  \\
			\cmidrule(lr){2-7}
			& \multirow{4}{*}{JVS}  & $\rho_l \equiv 2^{-1}$   & 3.03$\pm$0.052  & 2.07$\pm$0.025 & \textbf{3.11$\pm$0.044} & 3.08$\pm$0.045  \\
			&						 & $\rho_l \equiv 2^{-0.5}$ & 3.64$\pm$0.042  & 2.85$\pm$0.052 & \textbf{3.75$\pm$0.049} & 3.72$\pm$0.039  \\
			&						 & $\rho_l \equiv 2^{0.5}$  & \textbf{3.82$\pm$0.033}  & 3.19$\pm$0.058 & 3.60$\pm$0.053 & 3.63$\pm$0.054  \\
			&						 & $\rho_l \equiv 2^{1}$    & \textbf{3.08$\pm$0.040}  & 2.50$\pm$0.040 & 2.91$\pm$0.051 & 2.89$\pm$0.056  \\ 
			\bottomrule
		\end{tabular}
	\vspace{-0.6cm}
	\end{table*}

	Next, we compare the performance in terms of $f_0$ extrapolation. 1885 utterances from VCTK and 980 utterances from JVS were modified with the pitch scales $2^{-1}$, $2^{-0.5}$, $2^{0.5}$ and $2^{1}$, i.e., $\rho_l\equiv2^{-1}$, $\rho_l\equiv2^{-0.5}$, $\rho_l\equiv2^{0.5}$, and $\rho_l\equiv2^{1}$. The averaged quantized performance are listed in Table \ref{table_2}. Similarly, the frequency adaptation of QHM resulted in it achieving the best performance in terms of $f_0$ RMSE, whereas QHARMA-GAN achieves best V/UV rate. From the perspective of the quality of pitch-scaled speech, QHM showed the worst performance among the compared methods, since QHM does not model the spectral envelope and relies solely on a shape-invariant modification algorithm, making it difficult to accurately modify the speech signal. Moreover, when increasing $f_0$, the insufficient components in unvoiced parts further worsen the quality of pitch-scaled speech. On the other hand, QHARMA-GAN and WORLD achieved better results owing to their modeling of the spectral envelope. Notably, the MOS scores in Table \ref{table_2} reveal comparable performances: QHARMA-GAN, and even QHARMA-GAN-small, outperforms WORLD significantly in pitch lowering, while WORLD excels in pitch raising. A preliminary analysis suggests that this is because QHARMA-GAN relies on detected V/UV to selectively modify the speech. Unavoidable V/UV detection errors can reduce the number of components in which an unvoiced part is mistakenly classified as a voiced part, reducing the quality of the speech. Therefore, accurately estimating V/UV has become an urgent research topic for the future.

	\begin{table}[!t]\centering
		\caption{Results of RTFs computed on single Intel Xeon Gold 6230.}
		\vspace{-0.1cm}
		\setlength\tabcolsep{4pt} 
		\centering
		\label{table_3}
		\renewcommand{\arraystretch}{1.3} 
		\begin{tabular}{l|c|c|c}
			\toprule 
			RTF  $\downarrow$ & Analysis & Synthesis & Overall  \\
			\midrule
			WORLD               & 0.727 & 0.406 & 1.133        \\ \hline
			QHM                 &23.246 & 1.118 & 24.364      \\ \hline
			HiFi-GAN 	        &       &       & 0.153      \\ \hline
			Vocos 	            & 0.039 & \textbf{0.002} & \textbf{0.040}      \\ \hline 
			hn-NSF 	            &   &   & 0.192     \\ \hline 
			QHM-GAN 	        & 0.132 & 0.045 & 0.179   \\ \hline
			QHARMA-GAN 	        & 0.139 & 0.048 & 0.187      \\ \hline
			QHARMA-GAN-small    & \textbf{0.034} & 0.049 &0.084     \\
			\bottomrule
		\end{tabular}
		\vspace{-0.6cm}
	\end{table}

	\subsection{Evaluation of Model Efficiency}
	In this part, the efficiencies of all models are explored. It is quantized as RTF values in Table \ref{table_3}. We compare in detail the efficiencies of all methods in analysis and synthesis processes. In the analysis process, to ensure a fair comparison between WORLD, QHM, and the neural vocoders, we excluded the time consumption for $f_0$ detection in WORLD and QHM, because the computational process of the neural vocoders does not include $f_0$ detection, which is often a time-consuming step. From Table \ref{table_3}, it is obvious that the RTF of QHM is much higher than those of the other methods, because QHM needs to conduct least squares frame by frame. Likewise, the frame-by-frame estimation makes WORLD's analysis time-consuming. In contrast, neural vocoders are much faster during analysis because DNNs only require multiple simple computations, making them significantly quicker than conventional methods. The most notable example is Vocos, which achieves the best RTF of 0.040, much faster than HiFi-GAN and hn-NSF. This is attributed to the removal of transposed convolution, allowing it to quickly generate complex spectrograms and efficiently convert them into waveforms via iSTFT. In contrast, QHM-GAN achieves an RTF of 0.179 while QHARMA-GAN obtains RTF of 0.187. Furthermore, the simple structure of QHARMA-GAN-small accelerates the analysis process. In the synthesis part, QHARMA-GAN obtained a lower RTF (0.048) than WORLD and QHM, since the improved synthesis algorithm eliminates the need for frequency interpolation and instead directly interpolates the phase, reducing the computational cost. It is apparent that QHARMA-GAN is slower than HiFi-GAN, yet faster than hn-NSF. This is because hn-NSF works based on upsampled data and additionally incorporates a source module, which introduces the additional computation. While QHARMA-GAN improves analysis speed by avoiding upsampling during analysis, it brings additional computational burden during synthesis owing to the interpolations of both phase and amplitude. To address this, QHARMA-GAN-small has a simplified DNN structure, which improves the analysis speed and offsets the computational burden introduced by interpolation. In addition, further simplifying the DNN structure can accelerate the analysis part.

	\vspace{-0.3cm}
	\subsection{Evaluation of Generalization Ability}
	Generalization ability is of critical importance for neural vocoders. Most neural vocoders usually require large amounts of data for training to avoid overfitting, making them data-hungry in general, and HiFi-GAN is no exception. Therefore, in this part, we evaluate the generalization ability of several methods from two perspectives: out-of-distribution (OOD) evaluation and few-shot learning.

\begin{table}[t]\centering
	\caption{Results of objective evaluations for OOD evaluation. The UTMOS value of the ground truth was 2.42.}
	\setlength\tabcolsep{2pt} 
	\vspace{-0.1cm}
	\centering
	\label{table_ood}
	\renewcommand{\arraystretch}{1.3} 
	\begin{tabular}{l|c|c|c|c}
		\toprule 
		Metric & HiFi-GAN & Vocos & hn-NSF & QHARMA-GAN \\
		\midrule
		{V/UV rate {[}\%{]} $\downarrow$}    & 8& \textbf{7} & 8& \textbf{7}      \\ \hline
		{$f_0$ RMSE {[}Hz{]} $\downarrow$ }  &  0.16 & 0.28 & 0.16 & \textbf{0.11}    \\\hline
		{PESQ $\uparrow$}	  				 & 2.54 & \textbf{3.02} & 2.68 & 2.94     \\ \hline
		{MCD $\downarrow$}					 & 6.42 & \textbf{5.78} & 6.44 &5.96    \\ \hline
		{UTMOS $\uparrow$}   			     & 1.99 & 2.04 & 2.05 & \textbf{2.20} \\ 
		\bottomrule
	\end{tabular}
	\vspace{-0.6cm}
\end{table}

	\begin{table*}[!t]\centering
	\caption{Results of objective and subjective evaluations of small LJspeech. The MOS of the ground truth samples was $3.96\pm0.018$.}
	\vspace{-0.3cm}
	\setlength\tabcolsep{4pt} 
	\centering
	\label{table_4}
	\renewcommand{\arraystretch}{1.3} 
	\begin{tabular}{l|c|c|c|c|c}
		\toprule 
		Metric  & WORLD & QHM & HiFi-GAN & QHARMA-GAN & QHARMA-GAN-small \\
		\midrule
		{V/UV rate {[}\%{]} $\downarrow$}   & 11 & 10 & \textbf{9} & \textbf{9} & \textbf{9} \\   \hline
		{$f_0$ RMSE {[}Hz{]} $\downarrow$ } & 0.14 & \textbf{0.08} & 0.12 & 0.09 & 0.09 \\   \hline
		{PESQ $\uparrow$}	    		    & 2.64 & \textbf{3.63} & 3.13 & 3.18 & 3.18    \\ \hline
		{MOS $\uparrow$}          		    & 3.63$\pm$0.020 & 3.86$\pm$0.029 & 3.53$\pm$0.016 & 3.85$\pm$0.024 &   \textbf{3.90$\pm$0.022    }     \\		
		\bottomrule
	\end{tabular}
\end{table*}

	\subsubsection{Out-of-distribution Evaluation}
	We start from OOD evaluation, where we use the models trained on the JVS dataset (Japanese speech), including HiFi-GAN, Vocos, hn-NSF, and QHARMA-GAN, to synthesize the singing voices from OpenSinger \cite{opensinger}, which contains Mandarin and Cantonese songs performed by both male and female singers. To ensure gender diversity, we randomly selected 20 songs, comprising 727 utterances from both male and female singers to evaluate their performances. 
	
	Table \ref{table_ood} illustrates the averaged results of objective metrics, showing that QHARMA-GAN outperforms other methods in most metrics. This highlights the best generalization ability of QHARMA-GAN. Although Vocos slightly surpasses in terms of MCD and PESQ, its significantly higher $f_0$ RMSE reflects inferior pitch stability and degraded quality in singing voice synthesis. During the experiment, both HiFi-GAN and Vocos often generate distorted singing voices with unstable frequencies. Despite relying on predicted complex spectrograms for waveform generation, Vocos suffers from phase discontinuities across frames, resulting in perceptible frequency jitters in the synthesized singing voices. In contrast, hn-NSF absorbs the $f_0$ prior, enabling it to maintain a relatively stable $f_0$ in synthesized voice, even for such unseen data. Notably, QHARMA-GAN leverages an interpolation-based instantaneous phase reconstruction algorithm to maintain smooth and continuous frequency trajectories. Even when minor errors occur in phase estimation, the interpolation mechanism effectively compensates for such inaccuracies, further enhancing the naturalness and frequency coherence of the synthesized voices.
	
	Moreover, HiFi-GAN, Vocos, and hn-NSF fail to synthesize voices with extremely low or high $f_0$, especially for female soprano voices, due to the absence of such frequency ranges in the training corpus (JVS). Although hn-NSF leverages the $f_0$ prior, it still encounters difficulties in accurately modeling the resonance characteristics, thereby limiting its ability to generate high-quality voices under these challenging cases. In contrast, QHARMA-GAN, benefiting from the QHM-based processing pipeline, successfully synthesize singing voices with such extreme frequencies smoothly. These findings further confirm the superior generalization capability of QHARMA-GAN, particularly in extrapolating to unseen pitch conditions beyond the training distribution.

\begin{figure}[!t]\centering
		\vspace{-0.6cm}
		\includegraphics[width=8.9cm]{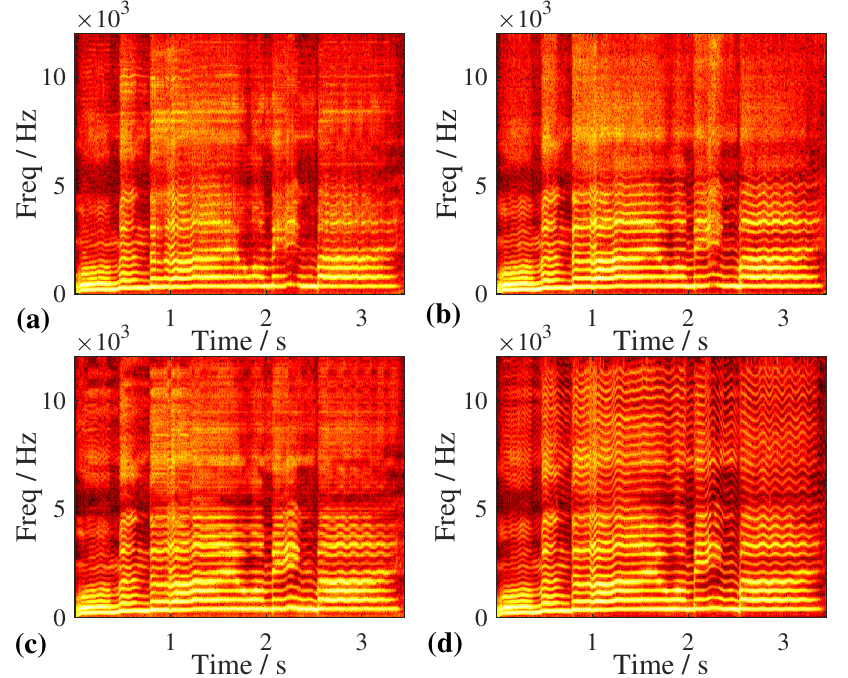}
		\caption{The spectrograms of the soprano voice generated by (a) HiFi-GAN, (b) Vocos, (c) hn-NSF, and (d) QHARMA-GAN.}
		\label{specs2}
		\vspace{-0.6cm}
	\end{figure}

	To illustrate this phenomenon, the spectrograms of a soprano voices generated by all methods are plotted in Fig. \ref{specs2}, showing that HiFi-GAN, Vocos, and hn-NSF synthesize only low-frequency harmonics, while failing to generate high-frequency harmonics, resulting in audible noise and distortion. In contrast, QHARMA-GAN clearly generates harmonics across the full frequency band, with smoother frequency curves, indicating that the generated voice from QHARMA-GAN is more natural and exhibits more stable pitch.

	\subsubsection{Few-shot Learning}
	To evaluate whether QHARMA-GAN can achieve good learning results with a smaller dataset, we trained both HiFi-GAN and QHARMA-GAN using a small portion of the single-speaker dataset LJspeech, i.e., 1419 utterances from LJspeech \cite{ljspeech17} split using a ratio of 919/250/250 for training, validation, and test sets. Table \ref{table_4} shows the results of metrics for all methods. It is apparent that QHM outperforms other methods in terms of objective indicators except the V/UV rate. QHARMA-GAN and HiFi-GAN obtained best V/UV rate since they can better model the unvoiced speech. The MOS scores showed that QHARMA-GAN can achieve a comparable performance to QHM in synthesis quality; even QHARMA-GAN-small obtained the highest score. Unfortunately, HiFi-GAN achieves the worst performance because of insufficient data for training, leading to overfitting. This undoubtedly exacerbates the frequency distortion in the synthesized speech. 
	
	The success of QHARMA-GAN trained by small dataset in synthesis quality implies that the framework with a combination of DNN and QHM exhibits strong generalization capabilities, alleviating the training pressure on the neural network. Humans are highly sensitive to the frequency of speech, in other words, any instability in frequency can significantly degrade speech quality. QHARMA-GAN generates speech by combining sinewaves, ensuring the smoothness of synthesized speech. Although phase delays of ARMA response adjust the frequencies, destroying the continuity of the phase of excitation signals, the cubic interpolation further consolidates the smoothness of both phase and frequency, thereby maintaining smooth frequencies for all sinewaves.

	\section{Conclusion}
	In this paper, to overcome the limitations of CSP vocoders, such as poor robustness and low synthesis quality, as well as the inability of neural vocoders to stably adjust $f_0$, we proposed a novel hybrid framework for neural vocoders, named QHARMA-GAN, which bridges the DNN and CSP through ARMA-based resonance characteristic modeling. The proposed framework consists of two parts: a DNN part for extracting ARMA coefficients for characterize resonance shape and a CSP part for synthesizing and modifying the speech using these coefficients. Firstly, the robustness of DNN was leveraged to overcome the low synthesis quality caused by the inaccurate resonance characteristic modeling of the CSP algorithm. Secondly, we proposed a novel CSP algorithm to quickly synthesize or modify speech waveforms by interpolating only amplitude and phase. Thus, our framework combines the strengths of both CSP and DNN and allows to rapidly leverage quasi-harmonics to reveal the structure of speech signals, enabling the model to synthesize speech and arbitrarily adjust the parameters, such as $f_0$, for modifying speech with high quality. The results of comparative experiments indicated that our proposed method can generate speech with frequency smoother than that of HiFi-GAN, while enabling a stable $f_0$ extrapolation of speech. Additionally, the non-upsampling structure and the simplification of the DNN part can improve the computation efficiency of the model, implying that QHARMA-GAN can achieve real-time processing, which can be embedded at the back end of some tasks, such as TTS and voice conversion. It is worth noting that, unlike vocoders that use only mel-spectrograms, QHARMA-GAN necessitates an additional module to predict $f_0$ when applied to TTS systems. However, such a framework offers better interpretability and enhanced controllability, holding promise for contributing to emotional TTS and prosody editing in the future.

	\bibliographystyle{IEEEtran}
	\bibliography{IEEEtranTSP}\
\end{document}